\documentclass[useAMS,onecolumn]{mn2e}
\usepackage{graphicx}
\usepackage{amsbsy}

\def\simlt{\lower.5ex\hbox{$\; \buildrel < \over \sim \;$}}
\def\simgt{\lower.5ex\hbox{$\; \buildrel > \over \sim \;$}}
\def\simpt{\lower.5ex\hbox{$\; \buildrel \propto \over \sim \;$}}

\def\kms{\mbox{ km s$^{-1}$}}
\def\mpc{\mbox{ Mpc}}

\def\msun{\mbox{ M}_\odot}

\def\mnras{MNRAS}
\def\apj{ApJ}
\def\apjl{ApJL}

\def\aap{AAP}

\def\prd{Phys.Rev.D}

\title{High-resolution imaging of the cosmic mass distribution from gravitational lensing of pregalactic HI}

\author[Metcalf \& White]{R. Benton Metcalf and S. D. M. White\\ Max Planck Institut f\"ur Astrophysics, Karl-Schwarzchild-Str. 1, 85741 Garching, Germany}

\begin{document} 

\maketitle

\begin{abstract}
Low-frequency radio observations of neutral hydrogen during and before the
epoch of cosmic reionization will provide $\sim 1000$ quasi-independent source
planes, each of precisely known redshift, if a resolution of $\sim 1$
arcminutes or better can be attained.  These planes can be used to reconstruct
the projected mass distribution of foreground material. Structure in these
source planes is linear and gaussian at high redshift ($30<z<300$) but is
nonlinear and nongaussian during re ionization. At both epochs, significant
power is expected down to sub-arcsecond scales.  We demonstrate that this
structure can, in principle, be used to make mass images with a formal
signal-to-noise per pixel exceeding 10, even for pixels as small as an
arc-second. With an ideal telescope, both resolution and signal-to-noise can
exceed those of even the most optimistic idealized mass maps from galaxy
lensing by more than an order of magnitude. Individual dark halos similar in
mass to that of the Milky Way could be imaged with high signal-to-noise out to
$z\sim 10$. Even with a much less ambitious telescope, a wide-area survey of
21 cm lensing would provide very sensitive constraints on cosmological
parameters, in particular on dark energy. These are up to 20 times tighter
than the constraints obtainable from comparably sized, very deep surveys of
galaxy lensing, although the best constraints come from combining data of the
two types. Any radio telescope capable of mapping the 21cm brightness
temperature with good frequency resolution ($\sim 0.05$~MHz) over a band of
width $\simgt 10$~MHz should be able to make mass maps of high quality.  The
planned Square Kilometer Array (SKA) may be able to map the mass with moderate
signal-to-noise down to arcminute scales, depending on the reionization history
of the universe and the ability to subtract foreground sources.
\end{abstract}

\section{introduction}

Dark matter appears to be the dominant component of all structures
larger than individual galaxies. In the standard paradigm its
gravitational effects drive the linear growth and the subsequent
nonlinear collapse of the fluctuations detected at $z\sim 1000$ in the
cosmic microwave background (CMB). Our inability to ``see'' the dark
matter, and so to image its distribution, has prevented a definitive
observational verification of this paradigm. Simulations of structure
formation predict all galaxies and galaxy clusters to sit within
extended dark halos with regular and well-specified structural
properties, but it has proved difficult to test these predictions
convincingly. As first demonstrated by \nocite{KS93}{Kaiser} \& {Squires} (1993) the distortion of
the images of distant objects caused by gravitational lensing can be
used to reconstruct an image of the foreground mass distribution. All
successful applications so far have used distant galaxies as the
sources. The resolution and signal-to-noise of the resulting maps are
fundamentally limited by the abundance and intrinsic ellipticity of these
sources. Even with deep satellite data the effective density of usable
galaxies does not exceed about 100 per sq.arcmin. For a map with 1
arcmin pixels this corresponds to a signal-to-noise per pixel (ratio
of {\it rms} expected physical fluctuation to {\it rms} noise
fluctuation) of about 0.75; for 10 arcmin pixels this ratio is about
2.5. As a result, only the centers of the most massive galaxy clusters
can be detected at high signal-to-noise in galaxy-based mass maps. In
this paper we show that much higher resolution and effective
signal-to-noise can, in principle, be achieved by using high redshift
neutral hydrogen as the source, rather than galaxies.

There has been a great deal of interest in the possibility of
observing the hyperfine transition of hydrogen (the 21~cm line) from
the intergalactic (or pregalactic) medium at high redshift
\nocite{astro-ph/0608032}(see Furlanetto, Oh \&  Briggs 2006, for an extensive review).  There are two
essentially disjoint epochs from which 21~cm radiation should be
observable.  At a redshift of $z\simeq 300$ neutral hydrogen (HI)
became thermally decoupled from the CMB.
The gas kinetic temperature then fell below the CMB temperature $T_r$
due to their different adiabatic cooling laws. For a while the spin
temperature $T_s$ remained coupled to the kinetic temperature by
atomic collisions, but at $z \sim 30$ the collision rate became so
low that the spin temperature decoupled from the kinetic temperature
and returned to equilibrium with the CMB.  During the period
$30<z<300$, $T_s$ was below $T_r$ and there was a net absorption of
CMB photons through the 21~cm line.  The observable quantity is the
brightness temperature $T_b = (T_s - T_r) (1-e^\tau)\simeq (T_s - T_r)
\tau $ which depends on the optical depth, $\tau$, which is in turn
proportional to the density of HI.  A map of $T_b$ on the sky and in
frequency would thus be a three dimensional map of the HI density,
which is directly proportional to the mass density at these redshifts.
The physics during this epoch of 21~cm absorption is simple, and
predictions within the standard cosmogony are straightforward and
robust.

The second epoch with observable 21~cm effects is considerably less
well characterized.  It is known that almost all intergalactic
hydrogen at $z<6.5$ is ionized.  It is believed that radiation from
the first generation of stars and/or quasars caused this reionization
between $z\sim 6.5$ and $z\sim 30$.  The latest CMB constraints give
$8.5<z_{reion}<22$ at 68\% confidence \nocite{wmap3year}(Spergel {et~al.} 2006).  A variety
of mechanisms will transfer energy from X-ray and/or Lyman-$\alpha$
radiation to the HI gas during reionization, thereby raising $T_s$
above the CMB temperature and making the 21~cm line visible in
emission. After reionization is complete, too little HI is left to be
observable.  The mean free path for X-rays through the neutral IGM at
$z<15$ can exceed the Hubble length, so the spin temperature for much
of the HI could have been raised uniformly before significant reionization
occurred.  Lyman-continuum radiation is expected to produce ionized
bubbles that expand until they overlap. Reionization finally completes
as the last interbubble clumps are evaporated. During this period,
Ly-$\alpha$ radiation passes freely through ionized regions but is
resonantly scattered in neutral regions, thereby raising their spin
temperature and producing 21 cm emission.  How rapid and inhomogeneous
this process was is highly uncertain and is likely to remain so until
it is directly measured.  It is also possible that shock heating of HI
gas during the collapse of pregalactic objects could raise $T_s$
enough for 21~cm emission to be visible before reionization begins
\nocite{2006ApJ...637L...1K}({Kuhlen}, {Madau} \&  {Montgomery} 2006).

Gravitational lensing distorts our image of the 21~cm emission and
absorption by moving the angular positions of points on the sky while
keeping the associated surface brightness (and thus brightness
temperature) unchanged.  For this reason a smooth background radiation
field is unaffected by lensing.  The observed map of brightness
temperature thus reflects both the intrinsic structure of the
fluctuations and the lensing distortions.  To separate the two, we use
the fact that in a given direction the intrinsic structure of maps at
sufficiently separated frequencies (hence redshifts) will be
statistically uncorrelated, while the foreground lensing distribution
will be the same.  Below we show how the maps can be combined so as to
average out the intrinsic temperature fluctuations while preserving
the lensing signal. In essence, the gradients of brightness
temperature maps at a set of sufficiently well-spaced frequencies are
independently and isotropically distributed in the absence of lensing,
but display a coherence which is a direct measure of the
lensing-induced shear when the foreground mass distribution is taken
into account.

Gravitational lensing of pregalactic 21 cm signals has previously been
considered by several authors. In particular, \nocite{ZandZ2006}{Zahn} \& {Zaldarriaga} (2006)
extended to 3-dimensions (angle on the sky + redshift of source) the
techniques developed by \nocite{2001PhRvD..64h3005H}{Hu} (2001) for detecting
lensing in the CMB, and they applied them to high-redshift 21~cm
emission.  In retrospect we find that the Fourier-space version of 
the method presented here is related to their method (see Appendices~\ref{app:lens-four-space} and 
\ref{app:conv-estim} for details) and that our method is related to one developed by \nocite{1999PhRvL..82.2636S}{Seljak} \& {Zaldarriaga} (1999) for detecting lensing in the CMB. \nocite{2004NewA....9..173C}{Cooray} (2004) had already
discussed applying the original 2-dimensional
\nocite{2001PhRvD..64h3005H}{Hu} (2001) method to the 21~cm absorption epoch, but
this misses the main advantage offered by the radio technique, namely
the large number of available quasi-independent source
planes. Finally, \nocite{2004NewA....9..417P}{Pen} (2004) discussed measuring
gravitational lensing effects in the 21~cm emission by looking for
anisotropic effects on the second order statistics of the brightness
fluctuations. This does not estimate the gravitational shear directly
by comparing maps at different frequencies in the same direction, and
so is much less sensitive than the approaches suggested here and by
\nocite{ZandZ2006}{Zahn} \& {Zaldarriaga} (2006).

The 21~cm emission/absorption has two major advantages over the CMB as
a background source for lensing studies.  Since lensing conserves
surface brightness, it can only redistribute structure that already
exists in the source.  The CMB has very little structure on the
angular scales where lensing is significant ($\simlt 1$~arcmin) so
that lensing effects are very weak.  The second advantage is that
the CMB provides only one temperature field on the sky while the 21~cm
emission/absorption provides many, all of which are lensed by the same
foreground mass distribution.  Although the CMB comes from higher redshift,
this is a relatively minor advantage since most of the structure
detected by lensing is at much smaller redshift than either source.

Our paper is organized as follows. In section~\ref{sec:shear_est} an estimator
for the gravitational shear is derived and in section~\ref{sec:noise} the
noise in that estimator is discussed and quantified using a particular model
for correlations in the 21~cm brightness temperature.  The expected lensing
signal and the size of objects that could be detected are calculated in
\S\ref{sec:sig_kappa}.  The prospects for measuring cosmological parameters
with 21~cm lensing are discussed in section~\ref{sec:darkenergy}.  The
observational prospects given currently planned telescope designs are
discussed in \S\ref{observations}.  In the appendices several technical issues
are addressed and alternative methods for measuring the lensing signal are
described.

\section{an estimator for the gravitational shear}
\label{sec:shear_est}

The observed deviation in the brightness temperature of the 21~cm emission at
a redshift $z$ (or equivalently frequency $\nu$) and a point on the sky,
$\vec{\theta}$ will be denoted $T(\vec{\theta},\nu)$.  We seek to construct a
statistic from this temperature that, when summed over frequency bands, $\nu$,
preserves the lensing signal while smoothing out the fluctuations in
$T(\vec{\theta},\nu)$.  A statistic will have these properties if it has the
same properties when averaged over an ensemble of temperature fields at a
fixed $\nu$ while keeping the lensing contribution fixed. All statistics that
are first-order in $T(\vec{\theta},\nu)$ vanish with this averaging because of
isotropy.

We will now show that it is possible to isolate the lensing
contribution in the second order statistics of the gradient of the
temperature field, $\vec{\nabla} T(\vec{\theta},\nu)$.  The small
angle, or ``flat sky'', approximation will be used throughout this paper and is well
justified for the angular scales that are considered.  The observed
temperature at a point on the sky, $\vec{\theta}$, is the source
temperature at $\vec{\theta}' = \vec{\theta} +
\vec{\alpha}(\vec{\theta},\nu)$ plus noise, where $\vec{\theta}'$ is
the position on the source plane (what the position would be in the
absence of lensing) and $\vec{\alpha}(\vec{\theta},\nu)$ is the
position shift caused by lensing (hereafter the deflection).  Thus 
the observed gradient of the temperature will be
\begin{eqnarray}\label{eq:divT}
\nabla_k T(\vec{\theta},\nu) = \left( \delta_{ki} + \alpha_{ki}
 (\vec{\theta},\nu) \right) \nabla'_i {\mathcal T}(\vec{\theta}',\nu)
 + N_k(\vec{\theta},\nu) ~~~,~~~ \alpha_{ki} (\vec{\theta},\nu) \equiv
 \frac{\partial \alpha_i(\theta,\nu)}{\partial \theta_k} ~,
\end{eqnarray}
where ${\mathcal T}(\vec{\theta},\nu)$ is the real, unlensed
brightness temperature and $\vec{N}(\vec{\theta},\nu)$ is the noise in
the measured gradient.  Repeated indices are summed over.  The square of
the magnitude of the observed gradient will be
\begin{eqnarray}\label{eq:dt2}
|\nabla T(\vec{\theta}) |^2 = |\nabla' {\mathcal T}(\vec{\theta}')|^2
+ \left( 2 \alpha_{ij}(\vec{\theta}) +
\alpha_{ik}(\vec{\theta})\alpha_{jk}(\vec{\theta}) \right)\nabla'_i
{\mathcal T}(\vec{\theta}')\nabla'_j {\mathcal T}(\vec{\theta}') \\ +
2\left(\delta_{ij} + \alpha_{ij}(\vec{\theta}) \right)
N_i(\vec{\theta})\nabla'_j {\mathcal T}(\vec{\theta}') +
|\vec{N}(\vec{\theta})|^2.  \nonumber
\end{eqnarray}
where the $\nu$'s have been left out for brevity.

The source emission, the deflection and the noise will all be
statistically independent so we can consider them separately.
Averaging over the source gives
\begin{eqnarray}
\left\langle \nabla'_i {\mathcal T}(\vec{\theta}',\nu) \right\rangle &=& 0~, \\
\left\langle \nabla'_i {\mathcal T}(\vec{\theta}',\nu)\nabla'_j {\mathcal T}(\vec{\theta}',\nu) \right\rangle & =& \frac{1}{2} \delta_{ij}\sigma_\nabla ^2(\nu)
\end{eqnarray}
where this defines $\sigma_\nabla ^2(\nu)$ and $\delta_{ij}$ is the
Kronecker delta. 
  
The distortion matrix $\alpha_{ij}$ can be decomposed into quantities
that are commonly used in lensing, the convergence $\kappa$, the shear
$\gamma$ and a rotation parameter $\beta$,
\begin{equation}
{\bf \alpha} = \left(
\begin{array}{ccc}
\kappa +\gamma_1  & \gamma_2  - \beta   \\
\gamma_2 + \beta &  \kappa-\gamma_1 
\end{array}
\right).
\end{equation}
Using this decomposition  we find
$\alpha_{ik}(\vec{\theta},\nu)\alpha_{jk}(\vec{\theta},\nu)$ to
correspond to the matrix
\begin{eqnarray}\label{eq:alpha2}
\left(\begin{array}{ccc}
\kappa^2+\gamma^2 + \beta^2 + 2(\gamma_1\kappa -\gamma_2 \beta)  & 2(\gamma_2\kappa+\gamma_1\beta)     \\
2(\gamma_2\kappa+\gamma_1\beta)   &  \kappa^2+\gamma^2 + \beta^2 -2(\gamma_1\kappa -\gamma_2 \beta)
\end{array}
\right).
\end{eqnarray}
The rotation term $\beta$ comes from coupling between different lens
planes and is second order in the surface density. It is expected to
be very small in nearly all cases so we will neglect it in what
follows, although its inclusion would be straightforward.  Because of
isotropy and the requirement that the usual angular size distance be
correct on average, we have $[ \alpha_{ij}(\vec{\theta},\nu) ]_\Omega=0$ where
$[...]_\Omega$ denotes an average over direction on the sky,
and
\begin{equation}
\sigma_\kappa^2(\nu) \equiv [ \gamma^2(\nu)]_\Omega = [ \kappa^2(\nu)]_\Omega\ ,
\end{equation}
where $\gamma^2=\gamma_1^2+\gamma_2^2$.  The second equality follows
from the deflection field being a potential field (i.e. assuming $\beta=0$).

We now construct three second-order quantities from the observed temperature gradient,
\begin{eqnarray}
\label{eq:Gamma1}
 \Gamma_1(\vec{\theta},\nu) & \equiv & \frac{1}{2} \left( \nabla_1 T(\vec{\theta},\nu) \nabla_1 T(\vec{\theta},\nu)   -  \nabla_2 T(\vec{\theta},\nu) \nabla_2 T(\vec{\theta},\nu)  \right) ,
\end{eqnarray}
\begin{eqnarray}
\label{eq:Gamma2}
 \Gamma_2(\vec{\theta},\nu) & \equiv &  \nabla_1 T(\vec{\theta},\nu) \nabla_2 T(\vec{\theta},\nu),
\end{eqnarray}
and
\begin{eqnarray}
\label{eq:Gamma3}
 \Gamma_3(\vec{\theta},\nu) & \equiv &  \frac{1}{2} \left| \nabla T(\vec{\theta},\nu) \right|^2,
\end{eqnarray}
where the indices on the gradient symbols refer to the two axes of the
chosen orthogonal coordinate system.  For a given direction (and hence
deflection field) the averages of these are
\begin{eqnarray}
\label{eq:aveGamma1}
\left\langle \Gamma_1(\vec{\theta},\nu) \right\rangle &=& \sigma_\nabla ^2(\nu)  \gamma_1(\vec{\theta},\nu)\left(1+\kappa(\vec{\theta},\nu)\right)  + \frac{1}{2}\left( \left\langle N_1(\nu)N_1(\nu)\right\rangle - \left\langle N_2(\nu)N_2(\nu)\right\rangle  \right) \\
\label{eq:aveGamma2}
\left\langle \Gamma_2(\vec{\theta},\nu)\right\rangle &  = & \sigma_\nabla ^2(\nu) \gamma_2(\vec{\theta},\nu)\left(1+\kappa(\vec{\theta},\nu)\right) + \left\langle N_1(\nu)N_2(\nu)\right\rangle \\
 \label{eq:avatpoint} 
\left\langle  \Gamma_3(\vec{\theta},\nu) \right\rangle & = & \frac{1}{2} \sigma_\nabla ^2(\nu) \left(  1+2\kappa(\vec{\theta},\nu) + \kappa(\vec{\theta},\nu)^2 + \gamma(\vec{\theta},\nu)^2  \right) + \frac{1}{2} \sigma_N^2(\nu) ,
\end{eqnarray}
where $\sigma_N^2(\nu)$ is the average of $|\vec{N}(\vec{\theta})|^2$
over random realizations of the noise.  If the noise is isotropic it
will drop out of both (\ref{eq:aveGamma1}) and (\ref{eq:aveGamma2}).
This can be seen by expressing the noise vector in terms of its
magnitude and polar angle and then requiring that the direction be
random.  However, in general the noise may not be isotropic so we
retain these terms.
To lowest order the first terms in the averages (\ref{eq:aveGamma1}) and
(\ref{eq:aveGamma2}) are proportional to the gravitational shear and
(\ref{eq:avatpoint}) is related to the convergence.  

The lensing signal from a single redshift slice will be dominated by
noise so we wish to add up frequency channels to reduce the noise. 
 The convergence and shear are slowly varying functions of $\nu$
at the high redshifts we are considering, so for now we will assume
$\gamma(\vec{\theta},\nu)$ to be independent of $\nu$ within the frequency band being used.  This
suggests estimating the shear at a point on the sky through
\begin{eqnarray}\label{shear_estimator}
\tilde{\gamma}_i(\vec{\theta}) =  \sum^{\nu_2}_{\nu=\nu_1} 
\omega_\nu \left\{ \Gamma_i(\vec{\theta},\nu) - \left[
\Gamma_i(\vec{\theta},\nu) \right]_\Omega \right\}
\end{eqnarray}
where the sum is over frequency channels.  The weights, $\omega_\nu$, are normalized so that the mean values are
\begin{eqnarray}
\left\langle \tilde{\gamma}_{1,2}(\vec{\theta}) \right\rangle =  \gamma_{1,2}(\vec{\theta}) \left(1+\kappa(\vec{\theta}) \right) \simeq \gamma_i(\vec{\theta}), \label{eq:ave_estimator}
\end{eqnarray}
and
\begin{eqnarray}
\left\langle \tilde{\gamma}_{3}(\vec{\theta}) \right\rangle =  \kappa(\vec{\theta}) + \frac{1}{2} (\kappa(\theta)^2+\gamma(\theta)^2 - 2\sigma^2_\kappa) \simeq \kappa(\vec{\theta}) .
\end{eqnarray}
The weights will be determined in the next section.  
Except along exceptional lines of sight (through the very centers of
galaxies and galaxy clusters) $\kappa(\vec{\theta})$ is much smaller
than one. As we show explicitly below, the variance $\sigma_\kappa^2$
is thus small, and (\ref{shear_estimator}) in effect provides an
unbiased map of $\vec{\gamma}(\vec{\theta})$ all the way back to the
beginning of structure formation.  We will sometimes refer to
$\tilde{\gamma}_i(\vec{\theta})$ as the shear estimators even though
the 3rd component is an estimator for the convergence, and
$\tilde{\gamma}_3(\vec{\theta})$ will sometimes be written as
$\tilde{\kappa}(\vec{\theta})$.

\section{noise levels}
\label{sec:noise}
\subsection{Instrumental, foreground and irreducible noise}

There will be a number of sources of noise in the estimators
(\ref{shear_estimator}). In particular, there will be noise from the
instrumentation, from terrestrial interference, and from incomplete
subtraction of galactic and extragalactic foreground emission.  This
noise is encapsulated in the $\vec{N}(\vec{\theta},\nu)$ vector field.
We will refer to these sources of noise collectively as {\it
foreground noise}.  It is expected that foreground emission will be
removed to high accuracy by using the fact that it varies slowly with
frequency, whereas the 21~cm emission/absorption signal (and
particularly the angular gradient of this signal) decorrelates for
even small separations along the
line-of-sight \nocite{2004ApJ...608..622Z,2005ApJ...625..575S}(see {Zaldarriaga}, {Furlanetto} \&  {Hernquist} 2004; {Santos}, {Cooray}, \&  {Knox} 2005).
The removal process could, however, leave noise with correlations in
both frequency and position on the sky.  For currently planned
generations of instruments, this residual is expected to be as small
as or smaller than the purely instrumental or thermal noise.  The lensing signal is also
coherent in frequency, but foreground subtraction will not effect it
because lensing is multiplicative while while the foregrounds are
additive, see equation~(\ref{eq:divT}).  Lensing does not cause
correlations between frequency channels, it causes spatial
correlations within a frequency channel that are the same as in the
other channels.

In addition to foreground noise there is noise from the randomness of
the $\nabla T(\vec{\theta},\nu)$ field itself. Clearly, this cannot be
reduced by any improvement in technology or foreground subtraction, so
we will refer to it as the {\it irreducible noise}.  It depends only
on the intrinsic correlations in the 21~cm signals and on the range of
frequency, or redshift, over which the signals are mapped.  We will
find that for any telescope which is able to {\it map} the 21 cm signals,
the total noise in the shear estimate will automatically be near the
irreducible value.  For this reason it is both a lower limit and a
good benchmark.

We must also differentiate between the noise per pixel and the noise in the average $\tilde{\gamma}(\vec{\theta})$
over a patch of sky which is larger than the pixel size.  By pixel we
mean the smallest resolvable region of the sky as set by the
telescope.  If the
angular correlations in the noise drop off more rapidly than the
correlations in the shear, then the signal-to-noise ratio will be
maximal on an angular scale that is larger than the pixel size.  We
will refer to a region of sky over which $\tilde{\gamma}(\vec{\theta})$
is averaged as a {\it patch} in the shear map.  A patch could be a
square region, a circular aperture, a gaussian smoothing window or any
other localized window.

The variances in the magnitude of our shear estimators,
(\ref{shear_estimator}), are given by
\begin{eqnarray}
\sigma^2_{\tilde{\gamma}}(\delta\Theta) & = &   \int\int d^2\theta d^2\theta' W(\vec{\theta};\delta\Theta)W(\vec{\theta'};\delta\Theta)
\sum_\nu \sum_{\nu'} ~\omega_\nu \omega_{\nu'} \sum_{i=1}^2\left\langle \Gamma_i(\nu,\vec{\theta}) \Gamma_i(\nu',\vec{\theta}') \right\rangle ~, \label{sig_g1} \\
\sigma^2_{\tilde{\kappa}}(\delta\Theta) &=&
\sigma^2_{\tilde{\gamma}_3}(\delta\Theta)  \\
&= & \int\int d^2\theta d^2\theta' W(\vec{\theta};\delta\Theta)W(\vec{\theta'};\delta\Theta)
\sum_\nu \sum_{\nu'} ~\omega_\nu \omega_{\nu'} \left[ \left\langle \Gamma_3(\nu,\vec{\theta}) \Gamma_3(\nu',\vec{\theta}') \right\rangle - \left\langle \Gamma_3(\nu)\right\rangle \left\langle\Gamma_3(\nu') \right\rangle \right]~, \label{sig_gam3}
\end{eqnarray}
where $W(\vec{\theta};\delta\Theta)$ is the window function defining
the patch (normalized to one when integrated over $\vec{\theta}$) and
$\delta\Theta$ is its characteristic angular scale. The noise in the magnitude of the shear per pixel
(i.e. in the original unsmoothed detection) is
$\sigma^2_{\tilde{\gamma}}(0)$ while the noise in the isotropic
estimator is $\sigma^2_{\tilde{\gamma}_3}(\delta\Theta)$.  
Note that the tildes are used to differentiate between noise in the
estimator, $\sigma_{\tilde{\gamma}}$ and the variance in the signal, $\sigma_{\gamma}$.
With some assumptions the correlation functions can be simplified
\begin{eqnarray}
 \sum_{i=1}^2\left\langle \Gamma_i(\nu,\vec{\theta}) \Gamma_i(\nu',\vec{\theta}') \right\rangle & = & 2\, \left\langle\Gamma_2(\nu,\vec{\theta})\Gamma_2(\nu',\vec{\theta}')\right\rangle \label{sig_g0}\ \\
 &=&2  \left\langle \nabla_1T(\nu,\vec{\theta})\nabla_2T(\nu,\vec{\theta})\nabla_1T(\nu',\vec{\theta}')\nabla_2T(\nu',\vec{\theta}') \right\rangle \label{sig_g2}\\
& = &2 \left\langle
  \nabla_1T(\nu,,\vec{\theta})\nabla_1T(\nu',\vec{\theta}')
  \right\rangle^2, \label{sig_g3}
\end{eqnarray}
and
\begin{eqnarray}
\left\langle \Gamma_3(\nu,\vec{\theta}) \Gamma_3(\nu',\vec{\theta}') \right\rangle - \left\langle \Gamma_3(\nu)\right\rangle \left\langle\Gamma_3(\nu') \right\rangle = \left\langle
  \nabla_1T(\nu,\vec{\theta})\nabla_1T(\nu',\vec{\theta}')  \right\rangle^2. \label{eq:sig_Gamm3}
\end{eqnarray}
  In (\ref{sig_g0}) we used the fact that rotational invariance requires
that the noise in both components of $\gamma(\vec{\theta})$ be the
same, so we choose to find the variance in the simpler $\Gamma_2(\nu)$
and double it to account for the other component.  To get from
(\ref{sig_g2}) to (\ref{sig_g3}) we have assumed that each component
of $\nabla T(\nu)$ is normally distributed so that the fourth moment
can be reduced to second moments. In addition, we assume that the
temperature field is isotropic so that there is no cross-correlation
between the components.  The same assumptions are used in expression~(\ref{eq:sig_Gamm3}).
 
Replacing the observed gradient in~(\ref{sig_g3}) with the true
temperature gradient plus noise gives the result
 \begin{eqnarray}
\sigma^2_{\tilde{\gamma}}(\delta\Theta) &=& 2 \sigma^2_{\tilde{\kappa}}(\delta\Theta) ~~~~~~~~~~~~~~~~~~~~~~~~~~~~~~~~~~~~~~\\  
&=&  \frac{1}{2} \sum_\nu \sum_{\nu'}~\omega_\nu \omega_{\nu'} {\mathcal A}(\nu,\nu',\delta\Theta)
 \label{sigma_g1}~,
\end{eqnarray}
where
\begin{eqnarray} \label{eq:Adef}
{\mathcal A}(\nu,\nu',\delta\Theta) \equiv
\int d^2\theta ~\overline{W}(\vec{\theta};\delta\Theta) \left[ \xi_\nabla (\nu,\nu',\theta)  + \xi_N(\nu,\nu',\theta) \right]^2 ,
\end{eqnarray}
\begin{eqnarray}
\overline{W}(\vec{\theta};\delta\Theta)  = \int d^2\theta'
~W(\vec{\theta}'+\vec{\theta}/2;\delta\Theta)W(\vec{\theta}'-\vec{\theta}/2;\delta\Theta) ,
\end{eqnarray}
and  
\begin{eqnarray}
\xi_\nabla (\nu,\nu',\theta=|\vec{\theta}'-\vec{\theta}''|) \equiv
\sum_{i=1}^2 \left\langle \nabla_i{\mathcal
T}(\nu,\vec{\theta}')\nabla_i{\mathcal T}(\nu',\vec{\theta}'')
\right\rangle  ~~~~~{\rm implying}~~~~~ \sigma^2_\nabla (\nu) = \xi_\nabla (\nu,\nu,0),
\end{eqnarray}
The correlation function $\xi_N(\nu,\nu',\theta)$ is similarly defined.  The pixel and frequency response functions 
are  included in ${\mathcal T}(\vec{\theta})$.

The optimal weights, $\omega_\nu$, can be calculated numerically, but a very good analytic approximation can be 
found by assuming that they vary slowly over the frequency range in which ${\mathcal T}(\nu)$ is correlated 
(${\mathcal A}(\nu,\nu') \simeq {\mathcal A}(\nu,\nu)$).  In this case $\omega_{\nu'}\simeq \omega_\nu$ in (\ref{sigma_g1}).  
Minimizing this subject to the constraint (\ref{eq:ave_estimator}) gives the weights
\begin{equation}\label{simple_weights}
\omega_\nu = \frac{\sigma^2_{\nabla}(\nu)}{\sum_{\nu'}{\mathcal A}(\nu,\nu')} \left( \sum_{\nu'} \frac{\left(\sigma^2_{\nabla}(\nu')\right)^2}{\sum_{\nu''}{\mathcal A}(\nu',\nu'')}  \right)^{-1}.
\end{equation}
Substituting this back into (\ref{sigma_g1}) gives the noise 
 \begin{eqnarray}\label{eq:final_sigma_gamma}
\sigma^2_{\tilde{\gamma}}(\delta\Theta) = \frac{1}{2}  \left( \sum_{\nu} \frac{\left(\sigma^2_{\nabla}(\nu)\right)^2}{\sum_{\nu'}{\mathcal A}(\nu,\nu')}  \right)^{-1}.
\end{eqnarray}

If the noise in the brightness temperature map is small compared to the fluctuations in the temperature itself, $\xi_N(\nu,\theta) < \xi_\nabla(\nu,\theta)$, which is a minimal requirement for mapping the
brightness temperature, then the foreground noise will drop out of (\ref{eq:Adef}) and the irreducible noise limit will be reached.  
To approach this noise level it is not necessary to eliminate all foreground noise. Thus a telescope designed to map
the brightness temperature will naturally achieve a noise level in
$\tilde{\gamma}(\vec{\theta})$ that is close to the irreducible value.
The correlations in frequency might be set by the bandwidth or by the intrinsic correlations in the brightness temperature.

It is often convenient to express the lensing noise (\ref{eq:final_sigma_gamma}) in terms of the
(cross-)power spectra of the brightness temperature
$\overline{C}^T_\ell(\nu,\nu')$ and the noise in the temperature
$\overline{C}^N_\ell(\nu,\nu')$.  This can be done by Fourier
transforming the temperature and gives
\begin{eqnarray} 
{\mathcal A}(\nu,\nu',\delta\Theta) = \int
\frac{d^2L}{(2\pi)^2}~|\tilde{W}(L,\delta\Theta)|^2 \int \frac{d^2\ell}{(2\pi)^2} ~ 
|\ell|^2|\vec{\ell}-\vec{L}|^2
\left(\overline{C}^T_\ell(\nu,\nu')+\overline{C}^N_\ell(\nu,\nu')\right)\left(\overline{C}^T_{|\ell+L|}(\nu,\nu')+\overline{C}^N_{|\ell+L|}(\nu,\nu')\right)
\end{eqnarray}
and
\begin{eqnarray}
\sigma^2_\nabla (\nu) = \int \frac{d^2\ell}{(2\pi)^2} ~ |\ell|^2 \overline{C}^T_\ell(\nu,\nu).
\end{eqnarray}
For a further discussion of calculating things in
Fourier-space see appendices~\ref{app:lens-four-space} and
\ref{app:conv-estim}.

To determine the possible capabilities of 21~cm lensing experiments we
now investigate the optimal case in which the irreducible limit is
reached with a bandwidth that is much smaller than the intrinsic
correlation length of the temperature.  In the limit of infinitely
narrow bandwidths the sums in eq.~(\ref{eq:final_sigma_gamma}) can be
converted into integrals. 
The function ${\mathcal A}(\nu,\nu',\vec{\theta})$ defines a volume
in frequency and angle within which the structure or noise is too
strongly correlated to contribute ``independent'' information to the
shear measurement.  A very useful approximation to this volume can be
found by calculating its characteristic length in frequency at
$\theta=0$ and its characteristic angular area at $\nu'=\nu$.  For the
temperature gradient alone these are
\begin{eqnarray}\label{eq:corr_length}
\Delta\nu_\nabla (\nu) \equiv \int_0^\infty d\nu'~\left( \frac{\xi_\nabla (\nu,\nu',0)}{\sigma^2_\nabla (\nu)} \right)^2 
\end{eqnarray}
and
\begin{eqnarray}\label{eq:corr_area}
\Delta\Omega_\nabla (\nu) \equiv \int d^2\theta~ \left( \frac{\xi_\nabla (\nu,\nu,\theta)}{\sigma^2_\nabla (\nu)} \right)^2. 
\end{eqnarray}
Analogous correlation lengths can be defined for the noise term and
for the cross-term.  Note that these correlation lengths are defined
with the correlation function squared, temperature to the fourth
power, which makes them significantly smaller than the usual
correlation lengths defined with the first power of the correlation
function.

When the patch size is near $\Delta\Omega_\nabla(\nu)$ there will be
only a few quasi-independent areas per patch.  To account for this it
is convenient to define the quantity
\begin{eqnarray}\label{eq:corr_frac}
{\mathcal N}_\nabla (\nu;\delta\Theta) \equiv   \int
d^2\theta~ \left( \frac{\xi_\nabla
(\nu,\nu,\theta)}{\sigma^2_\nabla (\nu)} \right)^2
\overline{W}(\vec{\theta};\delta\Theta).
\end{eqnarray}
This quantity is essentially the area of a correlated region divided
by the area of the patch. Two limiting cases are instructive.  For a
very small patch ${\mathcal N}_\nabla (\nu;\delta\Theta) \rightarrow
1$ and for a patch much larger than the intrinsic correlation length 
${\mathcal N}_\nabla (\nu;\delta\Theta) \rightarrow
\overline{W}(0;\delta\Theta) \Delta\Omega_\nabla (\nu)$
($=\Delta\Omega_\nabla (\nu)/(4\pi\delta\Theta^2)$ for a Gaussian
patch). One would like the data to be collected in
frequency channels with a width smaller than $\Delta\nu_\nabla(\nu)$,
otherwise the irreducible noise will be increased. Instrumental design
or foreground noise actually result in there being an optimal
bandwidth that is near $\Delta\nu_\nabla(\nu)$ as will be shown in
section~\ref{observations}.

Using the above definitions in (\ref{eq:final_sigma_gamma}) a simple
approximation for the irreducible
noise is found,
\begin{eqnarray}\label{eq:sig_irreducible}
\sigma^2_{\tilde{\gamma}}(\delta\Theta) \simeq \frac{1}{2} \left( \int^{\nu_2}_{\nu_1} d\nu~ \frac{1}{\Delta\nu_\nabla (\nu) {\mathcal N}_\nabla(\nu,\delta\Theta)} \right)^{-1}.
\end{eqnarray}
When the patch size is very close to the pixel
size the complete integrals in (\ref{eq:Adef}) must be carried out to
obtain an accurate result, but for the purposes of this section this
is not necessary.  The limit for small $\delta\Theta$ is the noise per
pixel.  It is easy to see from (\ref{eq:sig_irreducible}) and
(\ref{eq:corr_frac}) that the square of the irreducible noise is
essentially one over twice the number of correlated volumes in a patch.  

The above estimates assume that $\sigma^2_\nabla (\nu)$, the variance
in the intrinsic temperature within a frequency channel, can be
measured exactly so that the estimators $\tilde{\gamma}_i$ can be
normalized properly. This is normally a good approximation as we
now show.  The variance in the gradient can be found by averaging over position on the sky in the entire surveyed region.
Using (\ref{eq:dt2}) and dropping all
terms higher than second order in $\kappa$ and $\gamma$ results in
 \begin{eqnarray}
\sigma^2_K & \equiv &\left\langle \left[ \left( |\nabla T(\nu)|^2 \right)^2 \right]_\Omega \right\rangle  \\
& = &  2  \sum_\nu\sum_{\nu'}~\omega_\nu \omega_{\nu'} \int_\Omega d^2\theta  \left\{ 
\xi_\nabla (\nu,\nu',\theta)^2 \left(1+2\sigma_\kappa^2 + 2\xi_\kappa(\theta) \right) 
\right. \label{sigma_k1}\\
&& + 2 \xi_\nabla (\nu,\nu',\theta)  \xi_N(\nu,\nu',\theta)  \left( 1+\xi_\kappa(\theta) \right)  \label{sigma_k2} \\
&& \left. +  \xi_N (\nu,\nu',\theta)^2  \right\} \label{sigma_k3} 
 + \frac{4\sigma^2_\kappa}{\Omega} \int_\Omega d^2\theta ~ \frac{\xi_\kappa(\theta)}{\sigma^2_\kappa} \label{sigma_k4},
\end{eqnarray}
where the area of the surveyed region is $\Omega$.  It has been
assumed that $\sigma^2_N$ is determined by independent means to an
accuracy much better than the above.  The correlations in the lensing
convergence are relatively small ($\sigma^2_\kappa, \xi_\kappa(\theta)
\ll 1$) so the terms containing them on lines (\ref{sigma_k1}) and
(\ref{sigma_k2}) can be safely ignored.  The last term in
(\ref{sigma_k4}) expresses the uncertainty in the mismatch between the
average $\kappa$ (or $\gamma$) over the survey region and the true
average.

By comparing (\ref{sigma_k1})-(\ref{sigma_k4}) with
equation~(\ref{eq:Adef}) one can see that $\sigma^2_K \sim
\frac{4}{\Omega} ( \Delta\Omega_\nabla \sigma^2_{\tilde{\gamma}}(0) +
\Delta\Omega_\kappa \sigma^2_\kappa)$ where $\Delta\Omega_\kappa$ is
the area of sky over which the foreground convergence is
correlated. (Note that $\Delta\Omega_\kappa$ is defined with one power
of $\xi_\kappa(\theta)$ instead of two like $\Delta\Omega_\nabla$).
Both these correlated areas could effectively be as small as the pixel
if sparse pointings are used for normalizing
$\tilde{\gamma}(\vec{\theta})$.  If a survey covers just a few
independent regions and is capable of mapping the shear (so that
$\sigma_{\tilde{\gamma}} \simlt \sigma_\kappa$), then the noise in the
normalization of the shear map will be small, and
$\sigma_{\tilde{\gamma}}(\delta\Theta)$, as obtained above, can be
taken as the noise in the shear estimate.

\subsection{Correlations in the 21~cm emission}
\label{redshift_slices}

\begin{figure}
   \centering \includegraphics[width=3.5in]{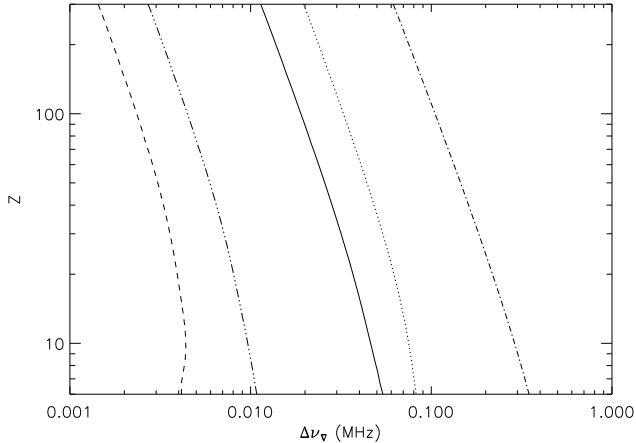}
   \caption{The frequency correlation length, $\Delta\nu_\nabla (z)$,
   defined in equation~(\ref{eq:corr_length}) as a function of
   redshift.  The power spectrum of 21~cm emission is taken to be the
   same as that of the dark matter, including linear velocity
   distortions and nonlinear structure formation.  The dotted-dashed
   curve is for a gaussian pixel of radius $\delta\theta=5$~arcmin, the
   dotted curve is for 1~arcmin, the solid curve is for 0.5~arcmin,
   the dot-dot-dot-dash curve is for 0.1~arcmin, and the dashed curve
   is for 0.05~arcmin (3~arcs). In the $\delta\theta =0.05$~arcmin
   case the decrease in the correlation length at small redshifts is
   caused by nonlinear structure formation.  This effect is present in
   the other cases but to a lesser extent.}  \label{fig:corr_length}
\end{figure}

\begin{figure}
   \centering \includegraphics[width=3.5in]{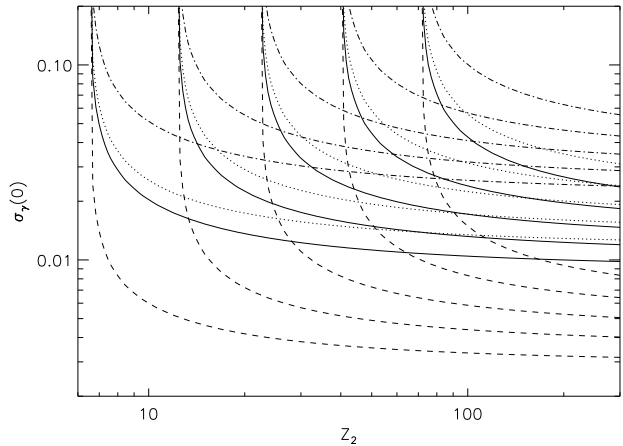}
   \caption{The expected irreducible noise in the shear measurement
   per pixel.  This is a plot of expression~(\ref{eq:sig_irreducible})
   with $\delta\Theta=0$ and assuming a CDM dark matter power spectrum
   for the 21~cm brightness temperature.  The dot-dash curves are for a
   pixel radius of $\delta\theta=5$~arcmin, the dotted curves are for
   1~arcmin, the solid curves are for 0.5~arcmin and the dashed curves
   are for 0.05~arcmin (3~arcs).  The upper limit of the redshift range
   used in the measurement is the abscissa, $z_2$.  For each pixel size
   the five curves are for different lower redshift limits.  They are
   from left to right (or down to up) $z_1=6.5$, 12, 22, 40 and 71.}
\label{fig:sig_g}
\end{figure}

\begin{figure}
   \centering \includegraphics[width=3.5in]{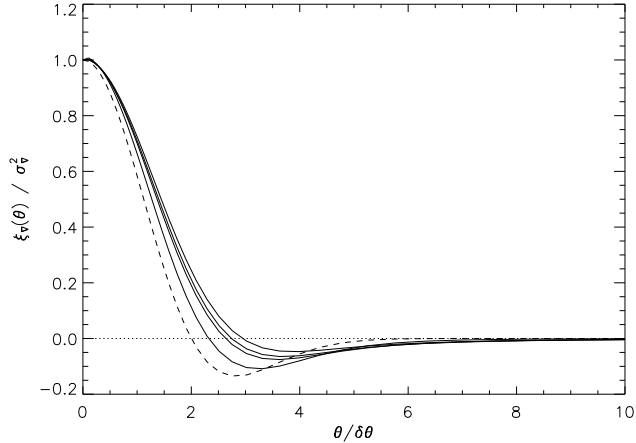}
   \caption{The angular correlation function at fixed frequency for
   different pixel sizes $\delta\theta$.  The pixel radii are 5, 1, 0.5
   and 0.05~arcmin.  Larger pixels have deeper minima.  These are for a
   source redshift of $z=18.9$, but all except the
   $\delta\theta=0.05$~arcmin case are very nearly independent of
   redshift.  The dashed curve is the result for the completely
   pixel-dominated Poisson process case.}  \label{fig:corrtheta}
\end{figure}

The irreducible noise in the shear map depends critically on the
number of statistically independent regions of 21~cm
emission/absorption along a single line-of-sight.  At the redshifts
where the 21~cm brightness temperature is significant the density of
the universe was dominated by ordinary matter so the comoving length
between two redshifts is well approximated by the flat
universe formula
\begin{equation}
l_{co} = \frac{2}{H_o\Omega_m^{1/2}(1+z_1)^{1/2}} \left[ 1 - \left(
\frac{1+z_1}{1+z_2} \right)^{1/2} \right]~.
\end{equation}
Between redshift 10 and 100, for example, $l_{co} = 2200 h^{-1}$~Mpc
for $\Omega_m = 0.3$, or $96.5 h^{-1}$~Mpc in proper distance.
Roughly speaking, the correlation length~(\ref{eq:corr_length}) is
$\sim 0.1 - 1$~Mpc (comoving) for a pixel size of 0.5~arcmin in radius
or smaller so we expect of order 1,000 independent samples between
these redshifts.  A more detailed calculation must take into account
the precise form of the correlations in the brightness temperature.

The irreducible noise is independent of the normalization of the
correlation function $\xi_\nabla(\nu,\nu',\theta)$ and thus will
depend only on the shape of the 3-dimensional correlation function or
power spectrum. During the early epoch of 21~cm absorption the
brightness temperature will be correlated in the same way as the dark
matter \nocite{2004PhRvL..92u1301L}({Loeb} \& {Zaldarriaga} 2004).  During reionization the
correlations could be very different.  One expectation is that
"bubbles" of ionized gas will form and expand until they merge.  The
size of the bubbles depends on the abundance and spatial distribution
of sources of ionizing radiation; AGN produce larger bubbles and stars
smaller bubbles.  These bubbles may or may not be smaller than the
pixel -- a 1~arcmin pixel has a comoving width of $1.9\,h^{-1}$~Mpc at
$z=10$.  In what follows we make the assumption that the brightness
temperature is proportional to the dark matter density even during
reionization.  We consider this conservative, because modifications to
the power spectrum during reionization are more likely to shorten the
correlation length (and so to reduce the noise) than to increase it.
The contribution of ionized bubbles will increase the correlations on
scales larger than the characteristic bubble size and suppress them somewhat on
scales smaller.  There have been
a number of attempts to model the fluctuations in the brightness
temperature during reionization
\nocite{2004ApJ...613..1,2004Natur.432..194W}(Furlanetto, Zaldarriaga \&  Hernquist 2004; {Wyithe} \& {Loeb} 2004).  We tried the simple
model of \nocite{2003ApJ...598..756S}{Santos} {et~al.} (2003) and find that it produces very
little difference in the irreducible noise for $\delta\theta =
0.5$~arcmin because the bubbles are significantly smaller than the
pixel sizes.  However, in the absence of either a complete theory of
reionization or direct observations, the form of the temperature
correlations remains a significant source of uncertainty in what
follows, especially for small pixel sizes.

The brightness temperature in direction $\vec{\theta}$ and at
frequency $\nu$ is given by
\begin{equation}
{\mathcal T}(\vec{\theta},\nu) = \int d^2\theta'\int dr ~{\mathcal
T}_{21}\left( \theta'_1D(\nu),\theta'_2D(\nu),r \right)
W(\vec{\theta}-\vec{\theta}') q_\nu\left(r-r(\nu)\right).
\end{equation}
where $q_\nu(r)$ is the response function of the telescope expressed
as a function of distance instead of frequency and $r(\nu)$ is the
comoving distance to the redshift from which the 21~cm line is
observed at frequency $\nu$.  Since peculiar velocities will
change the observed frequencies of the 21~cm line, $r(\nu)$ is
not actually the radial distance, but rather the redshift expressed as
a distance.

Using this we can find the correlation function between the gradient
of the temperature at different redshifts.  This can be done in
spherical coordinates, but it comes out much more simply in the small
angle approximation.  The bandwidth will initially be treated as
infinitely narrow, $q_\nu(r)=\delta(r(\nu)-r)$ (see
Appendix~\ref{app:lens-four-space} and \ref{app:conv-estim} for a treatment of finite bandwidths).  The result
expressed as an integral over Fourier-space is
\begin{eqnarray}
\xi_\nabla(\nu,\nu',\theta) & \equiv & \left\langle \nabla {\mathcal T}(\vec{\theta},\nu) \cdot \nabla {\mathcal T}(\vec{\theta}',\nu') \right\rangle \\ \nonumber
 & = & r(\nu)r(\nu')  \int_0^\infty \frac{d k}{(2\pi)^2}~  k^4 P_{21}\left( k,\overline{\nu} \right) ~{\mathcal W}(k,r(\nu),r(\nu'),\theta)
\end{eqnarray}
with
\begin{eqnarray}
{\mathcal W}(k,r(\nu),r(\nu'),\theta)&= & \int_{-1}^1 dx ~ (1+\beta x^2)^2(1-x^2)  \cos\left[k \Delta r(\nu,\nu') x \right] J_0\left[k\,\overline{r}(\nu,\nu')\theta \sqrt{1-x^2}\right]  \\
 \nonumber & & ~ \times  \tilde{W}\left[ k\, r(\nu) \sqrt{1-x^2} \right] \tilde{W}\left[ k\, r(\nu') \sqrt{1-x^2} \right]^\ast ~ , 
\end{eqnarray}
where $\Delta r(\nu,\nu')=r(\nu)-r(\nu')$,
$\overline{r}(\nu,\nu')=\left(r(\nu)+r(\nu')\right)/2$ and
$P_{21}(k,\overline{\nu})$ is the 3-dimensional power spectrum of the 21~cm
brightness temperature. It has been assumed that the power spectrum,
$P_{21}(k,\nu)$, does not change significantly over the range in $\nu$ where
there are significant correlations.  The linear redshift distortions
\nocite{1987MNRAS.227....1K}({Kaiser} 1987) are responsible for the $\beta$ term.  The
parameter $\beta$ is given by $\beta=\Omega_m(z)^{0.6}/b(z)^2$ to a very good
approximation, where $b(z)$ is the bias between the matter and the $T_{21}$
fluctuations and $\Omega_m(z)$ is the density of matter in units of the critical density at that time.  Here we have assumed that $P_{21}(k)=b^2 P_{\rm matter}(k)$ and
is thus proportional to the matter power spectrum.  In the calculations that
follow we take $b=1$, as expected at least during the early epoch of 21~cm
absorption.
  
With this result and with an assumed pixel profile the frequency correlation
length (\ref{eq:corr_length}), the angular correlation area
(\ref{eq:corr_area}) and the irreducible noise (\ref{eq:sig_irreducible}) can
all be calculated.  The nonlinear evolution of the power spectrum is of some significance for the smaller pixel widths considered here.  To account for this we use the \nocite{peac96}{Peacock} \& {Dodds} (1996) method to convert the linear power spectrum to a nonlinear one.

 Figure~\ref{fig:corr_length} shows $\Delta\nu_\nabla(z)$
as a function of redshift for a circular gaussian pixel with various radii
$\delta\theta$.  The decrease in $\Delta\nu_\nabla(z)$ with increasing
redshift is largely the result of a fixed comoving distance corresponding to a
smaller frequency interval at higher redshift.  The correlation length also
increases with increasing pixel size, but it is always less then 0.4~MHz, even when the
pixel is 5~arcmin in radius.

The irreducible noise per pixel, $\sigma^2_{\tilde{\gamma}}(0)$, is
shown in figure~\ref{fig:sig_g} for a few pixel sizes and ranges in
redshift.  It can be seen that smaller pixel sizes give {\it smaller}
irreducible noise per pixel.  It is not yet clear over what range in
$z$ the 21~cm emission/absorption will be detectable.  This depends on
the history of reionization, on the subtraction of foregrounds and on
telescope sensitivity.  If the reionization epoch lasts from $z\sim
10$ to 20 and this whole redshift range can be observed, then the
expected irreducible noise is $2\%$ for a $\delta\theta=0.5$~arcmin
pixel and $0.6\%$ for a 3 arcsec pixel.  The early epoch of 21~cm
absorption lasts from $z\sim 30 - 300$.  If this whole range could be
observed, then we expect $\sigma_{\tilde{\gamma}}(0)=1.7\%$ and
$0.6\%$ for the same pixel sizes.  It is possible that both epochs of
emission/absorption will someday be observable, reducing the noise still
further.

The angular correlation function at fixed frequency is shown in
figure~\ref{fig:corrtheta} for several different pixel sizes.
To a good approximation, the correlation function scales as
\begin{equation}\label{eq:fscaling}
\left( \frac{\xi_\nabla(\theta,\nu)}{\sigma^2_\nabla(\nu)} \right)^2 \simeq 4f_\epsilon f\left( \frac{\theta}{\delta\theta} \right),
\end{equation}
where $f_\epsilon$ is a constant.  The somewhat awkward normalization is
chosen so that $2\pi\int x f(x) dx =1$ and $f_\epsilon$ is unity to within a
few \% if the brightness temperature follows the CDM density field.  We retain
$f_\epsilon$ as a fudge factor which could differ significantly from one if
brightness temperature is not distributed like mass.  This approximate scaling
is a result of the power spectrum being almost scale-free on the relevant
scales.  It is a very useful approximation with important consequences,
because it means that the smaller the pixel, the lower the irreducible noise
for a fixed area on the sky.  The scaling can be understood by considering the
limiting case where the temperature is a Poisson process with correlations
only on scales much smaller than the pixel so that the pixelization
dominates the observed correlations. In this case
\begin{equation}
f(x) = \frac{1}{\pi} \left(1-\frac{x^2}{4}\right)^2 e^{-x^2/2}
\end{equation}
and $f_\epsilon = \pi/4$.  This limiting case is also shown in
figure~\ref{fig:corrtheta}.  The angular correlation is very nearly
frequency independent because comoving angular size distance is a slow
function of redshift at these high redshifts, and because the power
spectrum of temperature fluctuations does not change shape during
linear evolution.  There is some dependence on $\nu$ for the smallest
pixel radius ($\delta\theta=0.05$~arcmin) reflecting nonlinear
structure formation effects on these small scales ($\sim 100$~kpc) at
the lower redshifts.  The correlation area is a simple function of
$\delta\theta$, to a very good approximation $\Delta\Omega_\nabla
\simeq 4 f_\epsilon \delta\theta^2$.  Note that we use the radius of
the gaussian to characterize the pixel rather than its full width at
half maximum (fwhm).

Because of this simple scaling of correlated area with pixel size, a
simple expression for the irreducible noise per patch can be found
\begin{eqnarray}
\sigma^2_{\tilde{\gamma}}(\delta\Theta) & \simeq & \sigma^2_{\tilde{\gamma}}(0) {\mathcal N}_\nabla \\
& \simeq & \sigma^2_{\tilde{\gamma}}(0)\left\{
\begin{array}{ccl}
 \frac{1}{\pi} \left(\frac{\delta\theta}{\delta\Theta} \right)^2 f_\epsilon & , & \delta\Theta \gg \delta\theta \\
1 & , & \delta\Theta \ll \delta\theta 
\end{array}
\right.\label{eq:limits_sig}
\end{eqnarray}
To connect the two asymptotes the formulae~(\ref{eq:corr_frac}) and
(\ref{eq:sig_irreducible}) must be used.  The scaling as
$\delta\Theta^{-2}$ on large scales just reflects the fact that
correlations in temperature gradient are negligible on scales
significantly larger than the pixel. The prefactor might be different
if brightness temperature turns out not to be distributed like dark
matter density.  If the brightness temperature correlations have
strongly non-power-law behavior on the relevant scales, then
$f_\epsilon$ will show some dependence on the pixel size.  For example,
if the temperature distribution were smooth on small scales, then
making the pixel smaller would provide no further information and the
noise would not continue to decrease with pixel size.  As mentioned
before, the correlation length might also be smaller during
reionization, in which case $\sigma_{\tilde{\gamma}}(\delta\Theta)$
might also be smaller. This effect must be minor, however, since the
correlation length cannot be smaller than for the completely
pixel-dominated Poisson case and, as figure~\ref{fig:corrtheta} shows,
this is only slightly smaller than that of our standard model.

In Appendix~\ref{app:conv-estim} we present an alternative derivation of the noise in Fourier or visibility-space that agrees very well with the one given here.

\section{the expected signal and its connection to the distribution of matter}
\label{sec:sig_kappa}
\subsection{Signal-to-noise estimates}
\begin{figure}
   \centering
   \includegraphics[width=3.5in]{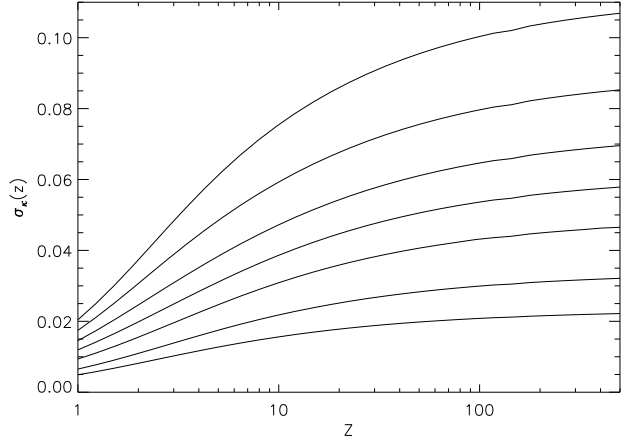}
   \caption{ The root mean square value of $\kappa$ smoothed with a
   gaussian window on the sky as a function of source redshift.  The
   curves from top to bottom are for patch sizes of 0.05, 0.2, 0.5, 1, 2,
   5 and 10~arcmin respectively.}
   \label{sig_kappa}
\end{figure}

\begin{figure}
   \centering \includegraphics[width=3.5in]{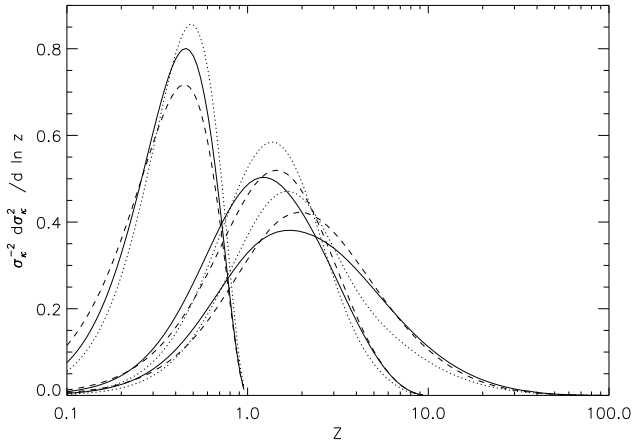}
   \caption{The contribution to $\sigma^2_\kappa$ from different
   redshifts, $\frac{1}{\sigma^2_\kappa} \frac{d}{d\ln z}\sigma^2_\kappa$.
   The curves end at the source redshifts $z=1$, 10 and 100.  The
   dotted curves are for a patch of radius 0.05~arcmin, the solid
   curves for 0.5~arcmin and the dashed curves for 5 arcmin.}
   \label{dsig_kappadz}
\end{figure}

\begin{figure}
   \centering
   \includegraphics[width=3.5in]{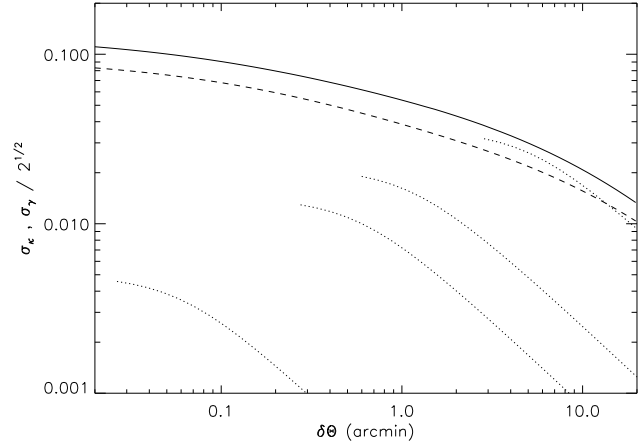}
   \caption{The root mean square convergence and the irreducible noise
     in the convergence
     ($\sigma_{\tilde{\kappa}}=\sigma_{\tilde{\gamma}}/\sqrt{2}$) 
   as a function of patch size.  The solid curve is
   $\sigma_\kappa(\delta\Theta)$ for a source redshift of $z=100$ and
   the dashed curve is the same for $z=10$.  The dotted curves are
   $\sigma_{\tilde{\kappa}}(\delta\Theta)$ for different pixel sizes --
   from top to bottom $\delta\theta = 5$, 1, 0.5 and 0.05~arcmin with
   a gaussian pixel.  The normalizations, $\sigma_{\tilde{\gamma}}(0)$
   are chosen to be representative, but will depend on the redshift
   range and the structure that exists in the 21~cm emission and
   absorption.  Guided by figure~\ref{fig:sig_g}, we have taken them
   to be 0.05, 0.03, 0.02 and 0.007 
   respectively.}  \label{sigKsigG}
\end{figure}

We now need to determine whether there will be enough signal on the appropriate angular
scales to produce a high fidelity map of the shear. This requires the noise to
be significantly lower than "typical" values of the shear. We quantify the
latter by calculating the root mean square value of the shear along random
lines of sight.

The distortion matrix introduced in section~\ref{sec:shear_est} can be written
in terms of derivatives of the Newtonian potential, $\phi(\vec{x})$, along the
light path.  To a good approximation the unperturbed light path can be used
(the first Born approximation) which results in
\begin{eqnarray} \label{eq:alphaofphi}
\alpha_{ij}(\vec{\theta},z) =   &   -2\int_0^{r(z)}dr'~ g\left(r(z),r'\right) \int d^2x_\perp W\left(\frac{\vec{x}_\perp}{D(r',0)} - \vec{\theta} ; \delta\Theta \right) \frac{\partial^2} {\partial x_i \partial x_j}  \phi(\vec{x}_\perp,r') 
\end{eqnarray}
with
\begin{eqnarray} \nonumber
g(r,r')\equiv \frac{D(r',0)D(r,r')}{D(r,0)}~,
\end{eqnarray}
where $r$ is the radial coordinate distance, $D(r,r')$ is the angular size
distance between the two coordinate distances and
$W(\vec{\theta};\delta\Theta)$ is still the angular window on the sky.
Sometimes the distances to the source redshift, to the lens redshift and
between them will be abbreviated as $D_s$, $D_l$ and $D_{ls}$,
respectively. The coordinate vector perpendicular to the line-of-sight is
$\vec{x}_\perp$.  The lensing convergence is
\begin{eqnarray}
\kappa(\vec{\theta},z) =   &   \frac{3}{4} H_o^2 \Omega_m \int_0^{r(z)}dr'~ (1+z') g\left(r(z),r' \right) \int d^2x_\perp W\left(\frac{\vec{x}_\perp}{D(r',0)} - \vec{\theta} ; \delta\Theta \right) \delta(\vec{x}_\perp,r')~,
\end{eqnarray}
where $\delta(\vec{x})$ is the fractional density fluctuation. 

To relate the variance in $\kappa$ to the power spectrum of matter
fluctuations it is easiest to use the Fourier space Limber's equation
\nocite{Kais92}({Kaiser} 1992) and then to transform back to angular space.  For a geometrically
flat universe the result is
\begin{eqnarray}
\sigma^2_{|\gamma|}=\sigma^2_\kappa(z) = \frac{9}{8\pi}\Omega_m^2H_o^3 \int_0^z \frac{dz'}{ E(z')} (1+z')^2\left(\frac{D(z',z)}{D(z)} \right)^2 \\ \nonumber
\times \int_0^\infty dl~ l~|\tilde{W}(\vec{l};\delta\Theta)|^2 P_\delta\left(\frac{l}{D(z')},z'\right)~,
\end{eqnarray}
where $E(z)=\sqrt{\Omega_m(1+z)^3 + \Omega_\Lambda}$ and $P_\delta(k,z)$ is
the 3D power spectrum of matter fluctuations at redshift $z$, and
$\tilde{W}(\vec{l};\delta\Theta)$ is the window in Fourier space.  The first
equality follows from the shear being a homogeneous potential field to first
order.  The window will be taken to be a gaussian to conform with our results
in section~\ref{redshift_slices}.

Figure~\ref{sig_kappa} shows $\sigma_\kappa(z)$ as a function of
source redshift for windows of different widths.  The expected
fluctuations in $\kappa$ are at the many percent level for redshifts
between 10 and 300 (for a 1~arcmin pixel 4\% to 6\%, for a 3~arcsec
pixel 7.5\% to 11\%).  Reducing the pixel size can increase the signal
substantially.  By comparing this figure with figure~\ref{fig:sig_g}
we see that for a pixel size of 1~arcmin or smaller and with a moderate
redshift range the irreducible noise per pixel is less than half the
expected signal. Figure~\ref{dsig_kappadz} shows which redshifts
contribute most to $\sigma^2_\kappa(z)$ for source redshifts of 1, 10
and 100.  It can be seen that structures above $z=2$ contribute
significantly in both 21~cm cases, whereas structure around $z\sim 0.5$
dominates in the galaxy lensing case.  If the shear could be measured
accurately using signals from both epochs of 21~cm emission/absorption,
one could expect to isolate the contribution from structure at $z\sim
10$, since this contributes significantly to the signal for source
redshift 100.  In these calculations the nonlinear power spectrum was
modeled using the \nocite{peac96}{Peacock} \& {Dodds} (1996) method with a normalization of
$\sigma_8=0.75$. Note that for the smaller pixels especially, the
distribution of $\kappa$ is strongly nongaussian, and the variance
plotted in fig.~\ref{sig_kappa} is substantially larger than a typical
fluctuation because of the long tail to high $\kappa$ values (see
\nocite{hilbert2007}{Hilbert} {et~al.} (2007).

Figure~\ref{sigKsigG} shows $\sigma_{|\gamma|}(\delta\Theta)$ and
$\sigma_{\tilde{\kappa}}(\delta\Theta)$ as functions of angular scale for
observations averaged over patches larger than the pixel.  The fluctuations in
shear drop off relatively slowly with increasing angular scale while at scales
much larger than the pixel size $\sigma_{\tilde{\kappa}}(\delta\Theta) \propto
\delta\Theta^{-1}$.  As a result even if the noise per pixel is comparable to
$\sigma_\kappa$ the shear can still be mapped with high signal-to-noise on
scales larger than the pixel.  With small noise per pixel, the surface density
averaged over large scales can be measured with high precision.  Note that the
normalizations of $\sigma_{\tilde{\kappa}}(\delta\Theta)$ in this plot depend
on the redshift range over which the 21~cm emission and absorption are
measured. We have chosen representative values as listed in the caption.

\subsection{Detection of individual objects}
\label{sec:detect_halos}
\begin{figure}
   \centering
   \includegraphics[width=3.5in]{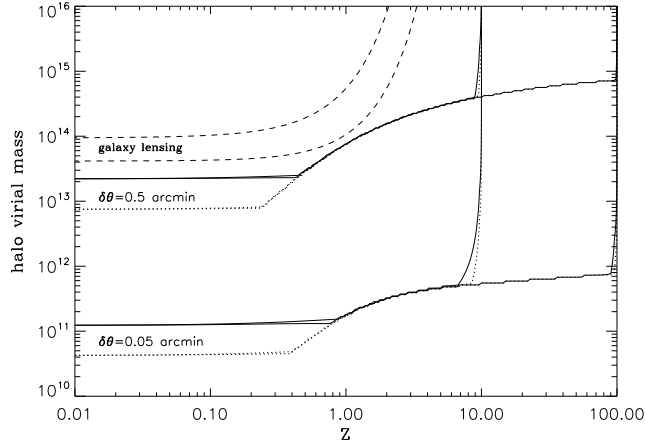}
   \caption{The mass detection limits for NFW halos as a function of
   redshift.  The dotted curves are $1\sigma$ and the solid curves are
   $2\sigma$ detections of the tangential shear in a circle centered
   on the halo.  The lowest set of four curves are for a pixel size of
   0.05~arcmin, $\sigma_{\tilde{\gamma}}=0.007$ and source redshifts
   of 10 and 100.  The middle set of four curves are for a pixel size
   of 0.5~arcmin and $\sigma_{\tilde{\gamma}}=0.02$.  The convex
   features in the curves at high redshift are a result of the
   requirement that the circle be larger then the pixel size.
The dashed curves are $2\sigma$ estimates for galaxy lensing surveys with the
   density of galaxies being, top, $35~\mbox{arcmin}^{-2}$ (typical of
   a ambitious ground based survey such as LSST or the DUNE satellite)
   and $100~\mbox{arcmin}^{-2}$ (perhaps achievable over a small
   region with a satellite such as SNAP). } \label{fig:detectionlimit}
\end{figure}

We have shown that 1$\sigma$ fluctuations in the convergence could be
detected with modest to good signal-to-noise (depending on pixel size) by a
21~cm experiment.  Another interesting question is what kind of objects would
be individually visible in a 21~cm shear map.  To answer this question let us
consider a circle of radius $\theta$ on the sky centered on a collapsed clump
or halo.  The average tangential shear on this circle is given by
\begin{equation}\label{eq:tanshear}
\overline{\gamma}_t(\theta) = \frac{4\pi G}{c^2} \frac{ D_{ls}}{D_l D_s} \left( \frac{M(\theta)}{\pi \theta^2} - \frac{1}{\theta}\frac{\partial M(\theta)}{\partial \theta} \right)~,
\end{equation}
where $M(\theta)$ is the mass within the circle.  The average
tangential shear within a disk can be found from this.  For halos with
an NFW profile \nocite{1997ApJ...490..493N}({Navarro}, {Frenk} \&  {White} 1997) we find, as a function of
virial mass and halo redshift, the radius where the signal-to-noise
for the average tangential shear is maximized.  The central density
and the scale-size are set according to the NFW prescription.  If the
disk is smaller than the pixel, the tangential component of the shear
will not be identifiable. Thus although these halos might cause a
significant feature in the shear map, we will not consider a halo
detected unless the signal-to-noise is above a 1 or 2-$\sigma$
threshold within a circle with a radius at least as large as the pixel
size.  The halo mass detection limit is plotted in
figure~\ref{fig:detectionlimit}.  With a 3~arcsec pixel this threshold
is below $10^{12}\msun$ almost all the way out to the redshift of the
21~cm emission/absorption and $< 2\times 10^{11}\msun$ below $z=1$.
This is smaller than the mass of the Milky Way halo today.  Note that
these are virial masses, not the mass enclosed within the circle. The
latter is the directly detected mass and can be significantly smaller.

Taking the average tangential shear over a disk is not the best method
for detecting halos.  One could do somewhat better by assuming a model
for their radial profiles and deriving an optimal weighting
function \nocite{1996MNRAS.283..837S}({Schneider} 1996). Here, however, we restrict
ourselves to the question of what objects would be clearly visible in
a shear map without any further special processing.

For comparison we calculate a similar mass limit for idealized future galaxy
lensing surveys.  In this case the noise in the average shear in a patch of
radius $\Theta$ is $\sigma^2 = \sigma^2_\epsilon/(\pi\Theta^2n_g)$ (half this
for just the tangential component) where $n_g$ is the angular number density
of background galaxies and $\sigma_\epsilon$ is the root-mean-squared
intrinsic ellipticity of those galaxies; we use the standard estimate
$\sigma_\epsilon = 0.3$.  The shear strength depends on the redshift
distribution of background galaxies with usable ellipticities.  Here we model
the redshift distribution as $\frac{dn_g}{dz} \propto z^2 e^{-(z/z_o)^{1.5}}$,
where $z_o$ is set by the desired median redshift.  The shear
(\ref{eq:tanshear}) must then be averaged over the portion of this
distribution that is at higher redshift than the lens plane. Halo detection
limits calculated in this way are also shown in
figure~\ref{fig:detectionlimit}.

A very deep space-based galaxy lensing survey might be competitive with a
$\sim 1$~arcmin pixel 21~cm lensing survey for detecting halos at $z<1$.  The
proposed satellite SNAP\footnote{snap.lbl.gov} is expected to survey $\sim
2\%$ of the sky with an expected galaxy density of $n_g\simeq100\mbox{
arcmin}^{-2}$ and a median redshift $z\sim 1.23$.  The
DUNE\footnote{www.dune-mission.net} satellite proposes surveying $\sim 50\%$
of the sky with $n_g\simeq 35\mbox{ arcmin}^{-2}$ and a median redshift of
$z\sim 0.9$. Several proposed ground based surveys --
LSST\footnote{www.lsst.org}, PanSTARRS\footnote{pan-stars.ifa.hawaii.edu},
VISTA\footnote{www.vista.ac.uk} would cover comparable areas to DUNE at
similar depth.  These are the two cases shown in
figure~\ref{fig:detectionlimit}.  Clearly higher redshifts will be accessible
with 21~cm lensing.  With a small pixel size, 21~cm lensing could detect all
Milky Way mass halos in the universe!  Based on the \nocite{2002MNRAS.329...61S}{Sheth} \& {Tormen} (2002)
halo mass function, for the same sky coverage $\sim 600$ times more objects
could be identified by such a survey than in a space-based galaxy shear map
with $n_g=100\mbox{ arcmin}^{-2}$ and $\sim 3,500$ times more than in a
ground-based galaxy shear survey with $n_g= 30 \mbox{ arcmin}^{-2}$.  Mass
maps of galaxy clusters could be made with arcsecond resolution and high
signal-to-noise, instead of with arcminute resolution and relatively low S/N
as is possible using galaxy lensing.  Galaxy halo studies, which now require
stacking thousands of galaxies to measure a single average shear profile,
could be carried out on individual galaxies.

\section{estimating cosmological parameters from the lensing power
  spectrum}
\label{sec:darkenergy}
\begin{figure}
   \centering
   \includegraphics[width=3in]{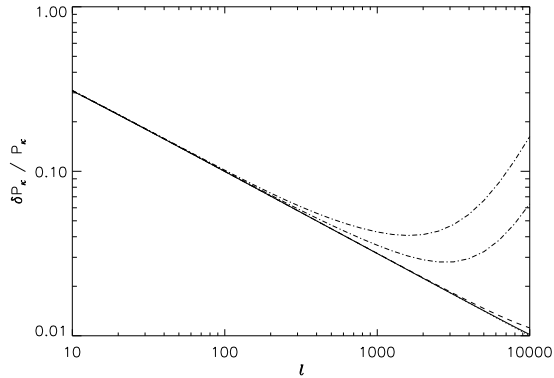} 
   \caption{The fractional error in the convergence power spectrum for a
   full-sky survey due to irreducible noise and cosmic variance.  The solid
   and dashed curves are for 21~cm lensing with sources at $z=100$ and $z=10$
   respectively, assuming a 0.5~arcmin pixel radius and
   $\sigma^2_{\tilde{\gamma}}(0)=0.03$.  The two dot-dashed curves are for
   galaxy lensing surveys with median source redshift $z=1$ and with 35 (upper)
   and 100 (lower) galaxies per square arcmin.  The dotted curve which is just
   visible in the lower right corner, but otherwise is covered by the solid
   curve is the cosmic variance limit.  For a smaller area survey these curves
   scale with the fraction of sky covered like $1/\sqrt{f_{\rm sky} }$ for
   modes that are smaller than the surveyed region. }
   \label{fig:powererror}
\end{figure}

As we have shown, high-resolution, high signal-to-noise shear maps
could be made using 21~cm lensing.  These maps will contain a wealth
of information which can be used not only to learn about structure
formation, but also to estimate cosmological parameters.  We will make
a preliminary foray into this latter topic in order to compare the
power of 21~cm lensing to that of galaxy lensing.  A useful study of
the the capability of planned galaxy lensing surveys for cosmological
parameter estimation has recently been published by
\nocite{amara&frig06}Amara \& Refregier (2006) and we will adopt their survey parameters in the following in order to facilitate
comparison between the two techniques.

Figures~\ref{sig_kappa} and \ref{dsig_kappadz} show clearly that the strength
of gravitational lensing depends on source redshift.  This suggests that
additional information may be extracted by comparing shear maps derived from
sources at different redshifts - either multiple 21~cm source planes, or
multiple galaxy source planes, or a combination of the two. Such weak lensing
tomography has already been proposed for galaxy lensing surveys as a method to
measure the evolution of structure and thereby to constrain the nature of dark
energy
\nocite{1999ApJ...514L..65H,2002PhRvD..66h3515H,2003MNRAS.343.1327H,2005PhRvD..72b3516C}({Hu} \& {Tegmark} 1999; {Hu} 2002; {Heavens} 2003; {Castro}, {Heavens} \&  {Kitching} 2005).
In this context 21~cm lensing has the potential advantages of superior
signal-to-noise, higher source redshift and better angular resolution.  On the
other hand, most models of dark energy affect structure formation and the
cosmic expansion rate primarily at $z\simlt 1$ where galaxy lensing tomography
is most sensitive.  As we show below, a combination of galaxy and 21~cm
lensing appears likely to constrain dark energy parameters most effectively.

For the purposes of cosmological parameter estimation it is convenient to work
in spherical harmonic or Fourier space. The cross-correlation between the
harmonic modes of two shear maps, corresponding to source planes at redshifts
$z_1$ and $z_2$, can be derived from equation~(\ref{eq:alphaofphi}) and is
directly related to the power spectrum of density fluctuations
\begin{equation}\label{eq:shearpower}
\begin{array}{c}
\left\langle \gamma_1(\ell,z_i) \gamma_1(\ell',z_j) \right\rangle \\
\left\langle \gamma_2(\ell,z_i) \gamma_2(\ell',z_j) \right\rangle \\
\left\langle \gamma_1(\ell,z_i) \gamma_2(\ell',z_j) \right\rangle 
\end{array}
=
\left(
\begin{array}{c}
\cos^2(2\theta_\ell) \\
\sin^2(2\theta_\ell) \\
\cos(2\theta_\ell)\sin(2\theta_\ell)
\end{array}
\right)\left\langle \kappa(\ell,z_i) \kappa(\ell',z_j) \right\rangle
\end{equation}
with
\begin{equation}
\left\langle \kappa(\ell,z_i) \kappa(\ell',z_j) \right\rangle = (2\pi)^2\delta^2(\ell - \ell') P^{ij}_\kappa(\ell)
\end{equation}
and
\begin{equation}
P^{ij}_\kappa(\ell)=\left[ \frac{3}{2} H_o^2 \Omega_m \right]^2 \int_0^{{\rm min}(r_1,r_2)} dr' \frac{g(r(z_i),r')g(r(z_j),r')}{D(r')^2} (1+z')^2 P_\delta\left(\frac{\ell}{D}\right)~,
\end{equation}
where $\ell$ and $\ell'$ are the multipole indices in the two maps.  Using
(\ref{eq:shearpower}) the shear (cross-)power spectrum is trivially converted into the
convergence (cross-)power spectrum.
 
The observed shear power spectrum of a lensing map contains a contribution
from the irreducible noise, but this term is absent in the cross-correlation
between maps for different source redshifts, since the noise fields in the two
source planes are then independent.  The power spectrum of the irreducible
noise can be found from the analysis of section~\ref{sec:noise}
\begin{eqnarray}
\left\langle \tilde{\gamma}(\ell) \tilde{\gamma}(\ell') \right\rangle
& = & (2\pi)^2 \delta((\ell - \ell') N_\kappa(\ell) \\
&\simeq &(2\pi)^2\delta^2(\ell - \ell') ~ \frac{1}{2(\nu_2-\nu_1)\overline{\Delta\nu_\nabla^{-1}}} \int d^2\theta \left( \frac{\xi_\nabla(\theta)}{\sigma^2_\nabla} \right)^2 e^{-i\vec{\ell} \cdot \vec{\theta}} \\
& \simeq  &(2\pi)^2\delta^2(\ell - \ell')  ~4
\sigma^2_{\tilde{\gamma}}(0) f_\epsilon\int d^2\theta~ f\left(
  \frac{\theta}{\delta\theta} \right) e^{-i\vec{\ell} \cdot
  \vec{\theta}}~
\end{eqnarray}
where the function $f(x)$ is defined by equation~(\ref{eq:fscaling}) and in
what follows it.  An alternative approach to calculating this noise
directly from visibility space is demonstrated in
appendix~\ref{app:conv-estim}.

The observed shear power spectrum including only the
irreducible noise will be
 \begin{eqnarray}
C_\kappa^{ij}(\ell) & = & P^{ij}_\kappa(\ell) \left|\tilde{W}(\ell)\right|^4  + N_\kappa(\ell) ~\delta_{ij}  \\
 & \simeq & P^{ij}_\kappa(\ell)e^{-2\delta\theta^2\ell^2} + 4  \sigma^2_{\tilde{\gamma}}(0) f_\epsilon\delta\theta^2 \tilde{f}\left(\ell \delta\theta \right) ~\delta_{ij}~. \label{eq:power_and_noise} \\ \label{eq:poissonPower}
& \sim & P^{ij}_\kappa(\ell)e^{-2\delta\theta^2\ell^2} + \pi \sigma^2_{\tilde{\gamma}}(0) \delta\theta^2 \left( 1 + \frac{(\delta\theta\ell)^4}{8} \right) e^{-\delta\theta^2\ell^2/2} ~\delta_{ij}  ~~~ \mbox{(resolution limited case)}
\end{eqnarray}
As can be seen in figure~\ref{fig:corrtheta}, the angular correlation function
$\xi_\nabla(\theta)$ has similar angular scale to the pixel.  As a result
$\tilde{f}(\ell\delta\theta)$ is close to unity for any $\ell \simlt 1/\delta\theta$ and it decreases rapidly for larger $\ell$.  We will
restrict ourselves to modes larger than the pixel (i.e. $\ell\delta\theta\ll
1$) in which case both $e^{-2\delta\theta^2\ell^2}$ and $\tilde{f}(\ell\delta\theta)$ will drop
out of $C_\kappa^{ij}(\ell)$.  Line (\ref{eq:poissonPower}) shows the result for the pixel dominated Poisson case discussed in section~\ref{redshift_slices}.

The shear power spectrum from galaxy lensing has the same form except there is no pixel.
It is often assumed that the noise in this case is dominated by the
intrinsic ellipticities of galaxies in which case the noise power
spectrum is $N_\kappa(\ell) = \sigma^2_\epsilon / n_{\rm g}$
\nocite{Kais92}({Kaiser} 1992).  In practice errors in the photometric redshifts of
the source galaxies are often important, but here we will assume an
ideal survey where these are not significant.

So far no assumption of Gaussian statistics in the shear field has been made in this section.
Although our estimator $\tilde{\gamma}(\vec{\theta})$ is not Gaussian, we have
shown that the correlation length of the irreducible noise should be close to
the pixel size.  The multipole moments for scales larger than the pixel size are
then sums of many independent variables and, by the central limit theorem, are
expected to be approximately normally distributed. The shear map itself will
also have substantial non-gaussianity caused by nonlinear structure, even for
 Gaussian initial density fluctuations.  However, on scales larger than
individual dark halos the shear map is expected to be close to Gaussian
because of contributions from many independent structures along the long
line-of-sight \nocite{2004MNRAS.348..897T}({Takada} \& {Jain} 2004).

For a Gaussian shear map the likelihood function factorizes by mode, making the
analysis much simpler.  The Fisher matrix in this case is
\begin{equation}\label{eq:fisher}
{\bf F}_{ab} = \frac{1}{2} \sum^{\ell_{\rm max}}_{\ell=\ell_{\rm min}} (2\ell +1) f_{\rm sky} {\rm tr}\left[ {\bf C}^{-1} {\bf C},_a {\bf C}^{-1} {\bf C},_b \right]~,
\end{equation}
(formula~(\ref{eq:fisher}) in appendix~\ref{app:param-estim-visib})
where the indices $a$ and $b$ refer to parameters $p_a$ and $p_b$ and 
$f_{\rm sky}$ is the fraction of the sky covered
\nocite{1999ApJ...522L..21H}({Hu} 1999).  The $f_{\rm sky}$ factor can be
interpreted as the result of limited resolution in visibility-space
because of the finite size of the radio telescope's pixel.  It is
assumed here that the coverage of the $u$-$v$ plane is complete
between $\ell_{\rm min}$ and $\ell_{\rm max}$ down to the resolution
of the telescope.
The smallest scale mode, $\ell_{\rm max}$, is
chosen so that the Gaussian assumption remains approximately valid.  The
minimum variance unbiased estimator of $p_a$ then has statistical
uncertainty $\sigma^2(p_a) = ({\bf F}^{-1})_{aa}$, so this quantity can be
used to indicate how well the parameter $p_a$ can be constrained.

Directly from (\ref{eq:fisher}) one finds that the power spectrum of
the fluctuations in $\kappa$ can be determined to accuracy
\begin{equation}\label{powererror}
 \Delta P_\kappa(\ell) = \sqrt{\frac{2}{(2\ell+1)f_{\rm sky}} }\left[ P_\kappa(\ell) + 4 \delta\theta^2\sigma^2_{\tilde{\gamma}}(0) f_\epsilon \tilde{f}\left(\ell \delta\theta \right) e^{2\delta\theta^2\ell^2}\right]
 \end{equation}
using only one epoch of 21~cm lensing.  This formula holds on scales between
that of the survey area, where windowing effects cause the noise to increase
sharply, and that of the pixel size.   Figure~\ref{fig:powererror}
shows the uncertainty in the $\kappa$ power spectrum given by this formula.
Not shown in the figure is the $\ell$-space resolution which is
$\Delta\ell \sim f_{\rm sky}$.  The errors in the power spectrum for
$\ell$ separated by less then $\Delta\ell$ will be correlated (see
appendix~\ref{app:param-estim-visib} for more details).  The cosmic variance (or sample variance for a partial sky survey) is likely to
dominate the uncertainty over all linear scales.  This illustrates a
fundamental limitation of measuring cosmological parameters from the
convergence power spectra and cross-power spectra.  Decreasing the
instrumental and/or irreducible noise does not provide any further information
about the ensemble power spectrum of $\kappa$ although it does provide more
information on the particular realization that we live in.  Modes with $\ell <
10^4$ will be cosmic variance limited if
$\sigma_{\tilde{\gamma}}(0)\delta\theta < 0.017$~arcmin for sources at $z=10$.
For modes with $\ell<10^3$ the same is true if
$\sigma_{\tilde{\gamma}}(0)\delta\theta < 0.12$~arcmin. When estimating
cosmological parameters there is no reason to decrease the noise below these
values, as long as the Gaussian assumption holds and one is only interested in
these modes.  Nevertheless, this does not mean that the cosmic variance limit
on the 3D power spectrum has been reached.  More information can be gained by
splitting the source redshift range up. This increases the noise for each
subrange, but accesses the additional tomographic information that is
averaged out when using the full redshift range to make a single shear map.

\begin{figure}
   \centering
   \includegraphics[width=5.5in]{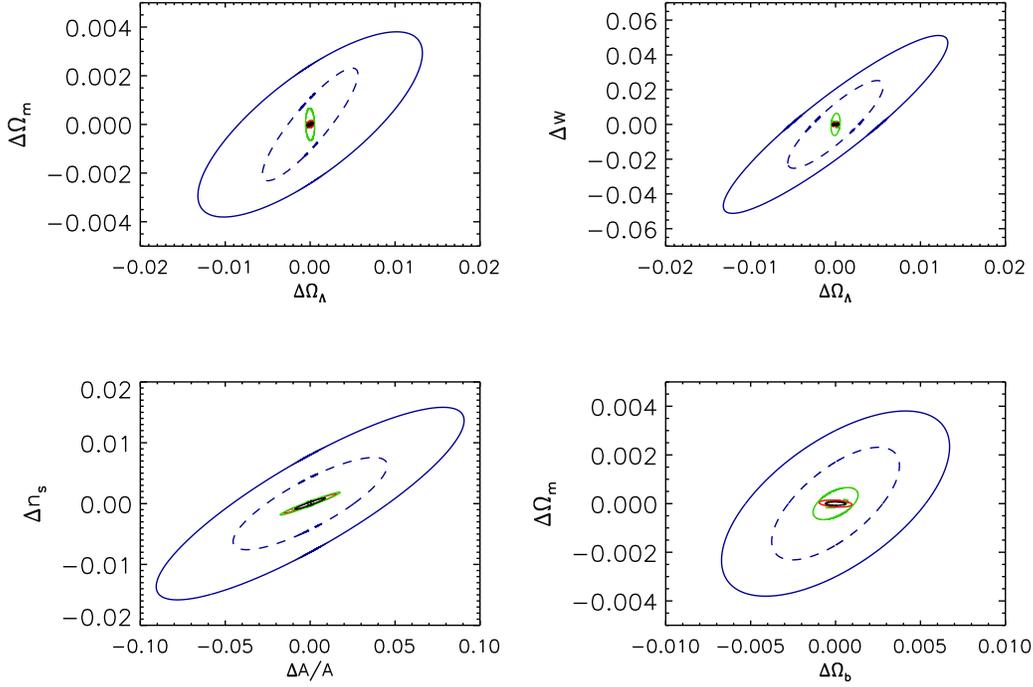} 
   \caption{Predicted error ellipses ($\chi^2=2.2789$ or 68\%
   probability) for six cosmological parameters. The fiducial model is
   $\Omega_m=0.3$, $\Omega_\Lambda=0.7$, $w=-1$, $n_s=1$,
   $\Omega_b=0.031$ and $\sigma_8 = 0.75$. The Hubble constant is
   fixed at $H_o=70\kms\mpc^{-1}$.  All constraints are obtained by
   marginalizing over the 4 parameters not shown in each plot.  The
   solid blue ellipses are for a full-sky galaxy survey with
   $n_g=35\mbox{ arcmin}^{-2}$, $\sigma_\epsilon =0.3$ and a median
   redshift of $z=0.9$.  The dashed blue ellipses are for a deeper
   survey with $n_g=100\mbox{ arcmin}^{-2}$, $\sigma_\epsilon =0.3$
   and a median redshift of $z=1.23$.  It would take the SNAP
   satellite roughly 30 years to complete a full-sky survey to the
   deeper depth. In all cases the galaxies are divided into three
   redshift bins as described in the text.  The solid green ellipses
   are for shear maps from 21~cm alone at redshifts $z=10$ and 15. The
   dashed green ellipses are for the ``optimal'' 21~cm case with shear
   maps constructed for $z=10$, 30 and 100. For these calculations we
   assume pixel radius $\delta\theta = 0.05$~arcmin and noise level
   $\sigma_{\tilde{\gamma}}(0)=0.02$, but the results are valid as
   long as $\sigma_{\tilde{\gamma}}(0)\delta\theta \simlt
   0.017$~arcmin and $\delta\theta < 0.5$~arcmin because in this case
   cosmic variance dominates the noise for all the $\ell$ values
   used. Modes $\ell=10$ to $\ell=10^4$ were used in deriving these
   constraints. The solid red ellipses are for the shallower galaxy
   survey combined with a 21~cm lensing survey at $z=10$ and 15.
   Finally, the solid black ellipse shows the ``optimum'' combination
   of the deeper galaxy lensing survey with 21~cm shear maps for
   $z=10$, 30 and 100. Figure~\ref{fig:covar_blowup} shows blow-ups of
   these plots so that the inner regions can be seen better.  The line
   types are summarized in Table~\ref{table:params}.}
   \label{fig:covar}
\end{figure}

\begin{figure}
   \centering
   \includegraphics[width=5.5in]{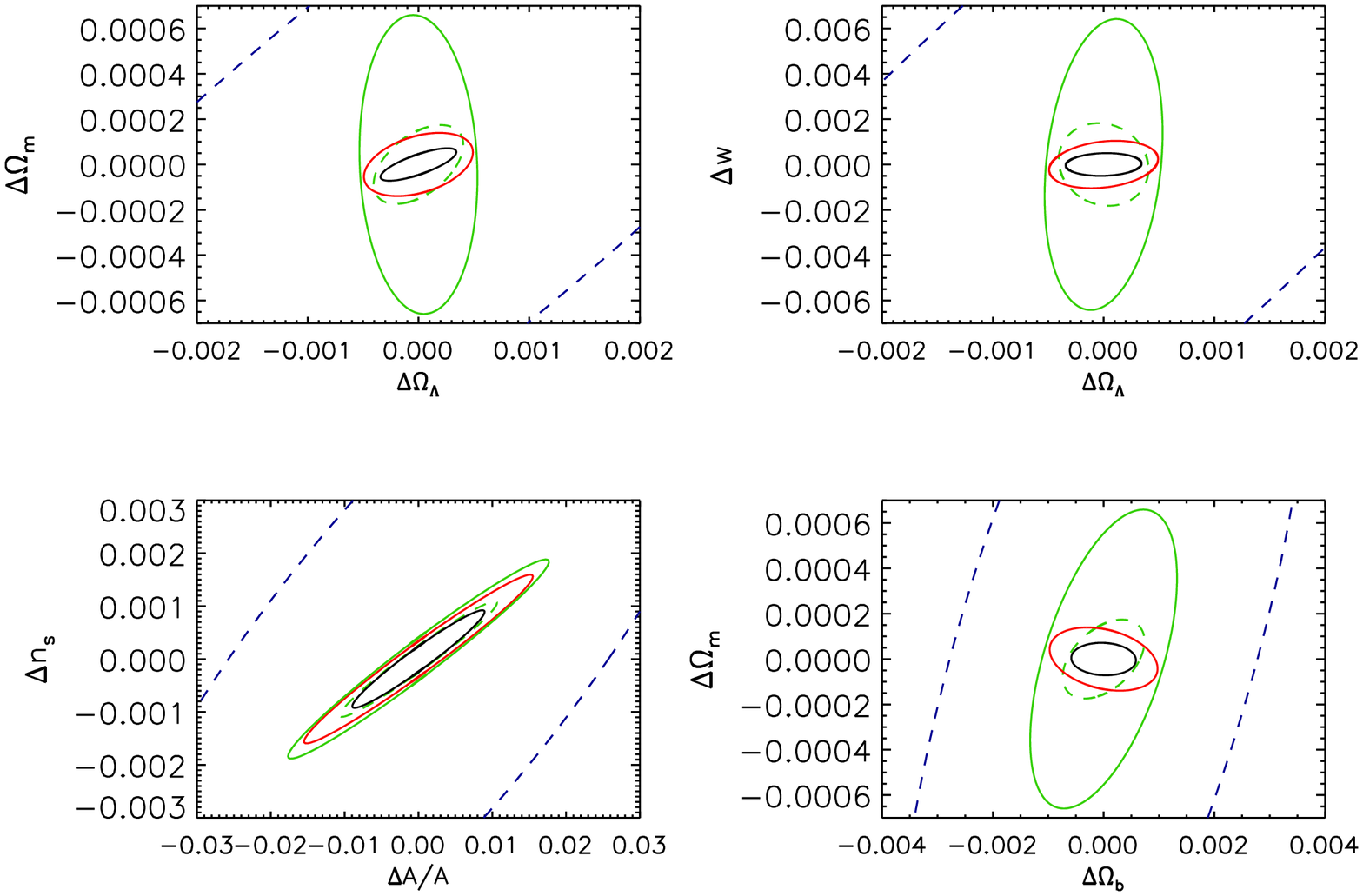} 
   \caption{Blow-ups of the plots in figure~\ref{fig:covar}.}
   \label{fig:covar_blowup}
\end{figure}

To proceed we must choose a cosmological parameter space for exploration, a
fiducial model to perturb around, and observational parameters for a set of
representative surveys.  For simplicity and for ease of comparison we follow
the galaxy survey parameters chosen by \nocite{amara&frig06}Amara \& Refregier (2006).  In the current standard paradigm, the apparently
accelerating expansion of the present universe is driven by dark energy, a
near-uniform and dominant component of the cosmic energy density with
effective equation of state $p= w \rho$ where $w < -1/3$
\nocite{2004ApJ...607..665R,AstierData,wmap3year}({Riess} {et~al.} 2004; {Astier} {et~al.} 2006; Spergel {et~al.} 2006).
Dark energy modifies the lensing signal due to cosmic structure in two ways.
It affects the angular size distance to a given redshift, which is given by the
expression
\begin{equation}
r(z) =\frac{1}{H_o}\int_0^z \frac{dz'}{E(z')}~,
\end{equation}
where
\begin{equation}
E(z) =  \left[  \Omega_m (1+z)^3 + \Omega_\Lambda \left( 1+z \right)^{3(1+w)}   + (1-\Omega_m-\Omega_\Lambda) (1+z)^2 \right]^{1/2}.
\end{equation}
Here $\Omega_\Lambda$ denotes the dark energy density today in units
of the critical density, and $w$ has been assumed to be constant.
The second effect of dark energy results from its influence on the linear evolution of density
fluctuations, which is given by
\begin{equation}
\frac{d^2\delta}{d\ln a^2} + \left( 2 + \frac{d \ln E(a)}{d\ln a} \right) \frac{d\delta}{d\ln a} -\frac{3}{2} \frac{\Omega_m}{a^3 E(a)^2} \delta =0.
\end{equation}
In addition to $\Omega_m$, $\Omega_\Lambda$ and $w$, we include in our
cosmological parameter set the logarithmic slope or spectral index of
the primordial power spectrum, $n_s$, the density of baryons,
$\Omega_b$, and the normalization of the power spectrum on large
scales $A$, which is proportional to $\sigma_8^2$.  The baryon
oscillations in the power spectrum are not calculated, so $\Omega_b$
only effects the overall shape.  Note that we do not restrict
ourselves to flat cosmologies, but we do fix the Hubble constant to
$H_o = 70$~km/s/Mpc, assuming this to be externally determined.
\begin{table}
\caption{Expected standard errors, $\sigma\times f_{\rm sky}^{1/2}$,
in cosmological parameter estimates based on various lensing
datasets.}
\begin{center}
\begin{tabular}{c|llllllll}
& galaxies & galaxies &  21 cm & 21 cm & galaxies shallow & galaxies shallow & galaxies deep\\
& shallow & deep & z=10, 15 & z=10, 30, 100 & 21 cm z=10 & 21 cm z=10, 15 & 21 cm z= 10, 30, 100\\
\hline
$\Delta\Omega_m$              & 0.0025 & 0.0015 &  0.0004 & 0.0001 & 0.0005 & $9.0\times 10^{-5}$ & $5.0\times 10^{-5} $ \\
$\Delta\Omega_\Lambda$ & 0.009    & 0.004   &  0.00035 & 0.0002 & 0.001  & 0.0003 & 0.0002\\
$\Delta w$                               & 0.03      & 0.02     &   0.004  & 0.001   & 0.002  & 0.0007    & 0.0003\\
$\Delta A$                               & 0.06 $\times A$ & 0.03 $\times A$ & 0.01$\times A$ & 0.007 $\times A$ & 0.02 $\times A$ & 0.01 $\times A$ & 0.006 $\times A$ \\
$\Delta n_s$                           & 0.01      & 0.005   & 0.001   & 0.0007  & 0.003 & 0.001   & 0.0006  \\
$\Delta\Omega_b$               & 0.004   & 0.002   &  0.0009 & 0.0005 & 0.002 & 0.0006 & 0.0004 \\
\begin{tabular*}{2.0cm}{l}
line type in \\
figures~\ref{fig:covar} \& \ref{fig:covar_blowup}
\end{tabular*}
& solid blue & dashed blue & solid green  & dashed green & - & solid red & solid black  \\
\end{tabular}
\end{center}
\label{table:params}
\end{table}

Figures~\ref{fig:covar} and \ref{fig:covar_blowup} show predicted
error ellipses for various pairs of our set of six cosmological
parameters and for various combinations of idealized 21~cm and galaxy
lensing surveys. Whenever calculations are done for 21~cm lensing at a
particular redshift the convergence is treated as if it where constant
over the redshift range used in estimating it.
For each plot we have marginalized over the remaining
four parameters of our model set. In Table~\ref{table:params} we give
corresponding $1\sigma$ uncertainties on individual
parameters after marginalizing over the other five dimensions of our
parameter space.  The galaxy redshift distributions assumed here are
the same as described at the end of section~\ref{sec:detect_halos}.
When the galaxies are binned into several redshift intervals, we
define these so as to obtain an equal number of galaxies in each bin.
We also assume the full sky to be surveyed in all cases; for partial
sky coverage the sizes of the uncertainties are approximately
increased by a factor of $f_{\rm sky}^{-0.5}$.
Apart from fixing the Hubble constant, no additional constraints from other
observations are included.  

In agreement with \nocite{amara&frig06}Amara \& Refregier (2006) we find that a ambitious galaxy
lensing survey could determine $\Omega_\Lambda$ and $n_s$ with an
accuracy of about $\sim 0.01$, $\Omega_b$ with an accuracy of about
0.004, $\Omega_m$ with an accuracy of about 0.0025 and $w$, with an
accuracy of about 0.03. An ideal survey going to a depth corresponding
to 100 source galaxies per square arcminute over the whole sky
(requiring about 30 years with the specifications of the SNAP
satellite) would reduce these uncertainties by about a factor of
two. Surveys covering only a fraction $f_{\rm sky}$ of the sky, would
have uncertainties increased approximately in proportion to $f_{\rm sky}^{-1/2}$.

While these numbers are impressive, shear maps derived from 21~cm
alone can provide considerably tighter constraints. All-sky maps
derived from the signal around $z=10$ and $z=15$ will be limited by
cosmic variance if their resolution and noise properties satisfy
$\sigma_{\tilde{\gamma}}(0)\delta\theta < 0.017$~arcmin, and will then
determine $\Omega_m$ to an accuracy of $4\times 10^{-4}$,
$\Omega_\Lambda$ to an accuracy of $3.5\times 10^{-4}$, and $w$ to an
accuracy of 0.004. An ideal survey with source planes around $z=10$,
30 and 100 would reduce these even further, with $\Delta\Omega_m\sim
10^{-4}$, $\Delta\Omega_\Lambda\sim 0.0002$, $\Delta\Omega_b\sim
0.0005$, and $\Delta w \sim 0.001$.

Some parameter constraints are substantially improved by combining
galaxy and 21~cm lensing, although most of the statistical weight
comes from the latter.  Thus combining the shallower galaxy survey
considered above with the 21~cm survey at $z=10$ and 15, one finds
that $\Omega_\Lambda$, $A$, $n_s$ and $\Omega_b$ are constrained about
as well as by the 21~cm alone, while $w$ is constrained almost six
times better and $\Omega_m$ four times better.  Constraints on dark
energy parameters are improved by including the galaxy lensing because
dark energy primarily affects structure evolution at $z<1$.  On the
other hand, galaxy lensing alone gives comparatively poor constraints
on these parameters unless a prior constraint on $\Omega_m$ is
included.  For parameters that affect only the matter power spectrum
(e.g. $n_s$) 21~cm lensing has a larger comparative advantage.  Of
course it is not a question of one or the other.  Clearly it is worth
doing both galaxy and 21~cm lensing surveys to maximize the
information gained and to spread the risk from unanticipated
systematics.

It should be emphasized that this analysis does not exhaust the
potential for constraining cosmological parameters using 21~cm or
galaxy lensing.  The dark energy model used here is overly simplified
and may be unrealistic; some more physically based models imply
appreciable effects at redshifts well beyond unity and so may be
particularly well constrained by 21~cm surveys
\nocite{2003ApJ...591L..75C,2001PhRvD..64l3520D}({Caldwell} {et~al.} 2003; {Doran}, {Schwindt} \&  {Wetterich} 2001). Other datasets,
notably CMB observations and supernova surveys, constrain cosmological
parameters in different ways than gravitational lensing, and will be
much improved by the time surveys of the type discussed in this
section are completed. Combining results from all these sources will
give stringent tests for the presence of systematics and will provide
tighter and more robust final constraints if overall consistency is
found.  Our knowledge of many cosmological parameters is limited by
degeneracies which are drastically reduced when different types of
observation are combined in this way.  

\section{observations}
\label{observations}

So far we have considered idealized observations where the irreducible noise
dominates and the bandwidth is smaller than the intrinsic correlation length
of the brightness temperature.  This will be the best any experiment can do
and, as we showed, will be reached when the noise in the temperature map is
small compared to the temperature fluctuations in each frequency channel.
This irreducible noise depends only on the shape of the temperature
correlation function.  Realistic observations, at least in the near future,
will have foreground noise levels that are comparable to or larger than the
intrinsic fluctuations in the brightness temperature.  In this case the noise
in the lensing map will depend more sensitively on the parameters of the
telescope and on the level and statistical properties of the brightness
temperature fluctuations. We now discuss these factors in more detail.

The observations will be carried out with radio interferometers and thus in
visibility space. As a result, when calculating the performance of telescopes
it is easier to work in Fourier-space.  For this section we adopt the
formalism of appendix~\ref{app:conv-estim} for convenience.
Equations~(\ref{sigma_from_l}) and (\ref{N_L_nocorr}) give the noise in the
$\kappa$ estimate as a function of the power spectrum of foreground noise,
$C_\ell^N(\nu)$, and the power spectrum of the brightness temperature,
$\overline{C}_\ell(\nu)$.

The noise in each visibility will have a thermal component and a component
from imperfect foreground subtraction.  We will model only the thermal
component.  If the telescopes in the array are uniformly distributed on the
ground the average integration time for each baseline will be the same and the
power spectrum of the noise will be
\begin{eqnarray}
 C_\ell^N = \frac{2\pi}{ \Delta\nu t_o} \left(  \frac{T_{\rm sys} \lambda}{ f_{\rm cover} D_{\rm tel}} \right)^2 = \frac{(2\pi)^3 T_{\rm sys}^2}{\Delta\nu t_o f_{\rm cover}^2 \ell_{\rm max}(\nu)^2} 
\end{eqnarray}
\nocite{2004ApJ...608..622Z,2005ApJ...619..678M,2006ApJ...653..815M}({Zaldarriaga} {et~al.} 2004; {Morales} 2005; {McQuinn} {et~al.} 2006) where
$T_{\rm sys}$ is the system temperature, $\Delta\nu$ is the bandwidth, $t_o$
is the total observation time, $D_{\rm tel}$ is the diameter of the array and
$\ell_{\rm max}(\lambda)=2\pi D_{\rm tel}/\lambda$ is the highest multipole
that can be measure by the array as set by the largest baselines.  $f_{\rm
cover}$ is the total collecting area of the telescopes divided by $\pi (D_{\rm
tel}/2)^2$, the covering fraction.  Other telescope configurations are
possible which would make the noise unequally distributed in $\ell$, but we
will consider only this uniform configuration here.  For our calculations we
will use $T_{\rm sys} = 200$~K.

There are several relevant telescopes proposed or under construction.  The 21
Centimeter Array (21CMA, formerly known as PAST) has $f_{\rm cover} \sim 0.1$
and $\ell_{\rm max} \sim 10^3$ giving it a resolution of about 10~arcmin.  The
Mileura Widefield Array (MWA) Low Frequency Demonstrator
(LFD)\footnote{ttp://www.haystack.mit.edu/ast/arrays/mwa/} will operate in the
80-300~MHz range with $D_{\rm tel}\simeq 1.5$~km and $f_{\rm cover} \sim 0.1$.
For LOFAR (Low Frequency Array)\footnote{www.lofar.org} the core array will
have $f_{\rm cover} \sim 0.016$ and $D_{\rm tel} \sim 2$~km.  LOFAR's extended
baselines, out to 350~km and possibly larger, are not expected to be useful
for high redshift 21~cm observations because of the small $f_{\rm cover}$ of
the extended array, although they will be used in foreground subtraction.  It
is anticipated that it will be able to detect 21~cm emission out to a redshift
of $z\simeq 11.5$, but sensitivity limitations will make mapping very
difficult.  Plans for SKA (Square Kilometer
Array)\footnote{www.skatelescope.org/} have not been finalized, but it is
expected to have $f_{\rm cover} \sim 0.02$ out to a diameter of $\sim 6$~km
($\ell_{\rm max} \sim 10^4$) and sparse coverage extending out to
1,000-3,000~km.  The lowest frequency currently anticipated is $\sim 100$~MHz
which corresponds to $z\sim 13$.  It is anticipated that the core will be able
to map the 21~cm emission with a resolution of $\delta\theta = \Delta\theta/2
\sim 0.5\mbox{ arcmin}$. For reference what we call the pixel-width is given
by $\Delta\theta = 2\delta\theta \sim \pi/\ell_{\rm max} $ or $1.08\times
10^4/\ell_{\rm max}$~arcmin.  One arcminute (fwhm) corresponds to baselines of
7.9~km and 73~km at redshifts of 10 and 100 respectively.  For our calculation
we will concentrate only on an SKA-like array with $D_{\rm tel}=6$~km and a redshift 
range out to $z=13$ since the smaller planned telescopes will not be capable of mapping mass at high fidelity.

The fluctuations in the brightness temperature depends on the spin
temperature, the ionization and the density of HI through
\begin{eqnarray}
\delta T_b \simeq 24  (1+\delta_b) x_{\rm H}\left( \frac{T_s- T_{\rm CMB}}{T_s}\right)\left(\frac{\Omega_b h^2}{0.02}\right) \left( \frac{0.15}{\Omega_m h^2} \frac{1+z}{10} \right)^{1/2} \mbox{ mK}
\end{eqnarray}
\nocite{1959ApJ...129..536F,1997ApJ...475..429M}({Field} 1959; {Madau}, {Meiksin} \&  {Rees} 1997).  As is commonly done, we will
assume that the spin temperature is much greater than the CMB temperature.
This leaves fluctuations in the ionization fraction, $x_H$, and the baryon
density $\delta_b = (\rho_b - \overline{\rho}_b) /\overline{\rho}_b$ as the
sources of fluctuations.  We will make the simplifying assumption that $x_{\rm
H}=1$ until the universe is very rapidly and uniformly reionized.
Realistically, the reionization process will be inhomogeneous and may extend
over a significant redshift range.  This will increase
$\overline{C}_\ell(\nu)$ by perhaps a factor of 10 on scales larger the
characteristic size of the ionized bubbles and thus might be expected to
reduce the noise in $\tilde{\kappa}$ significantly.  However, we have derived
the noise in the lensing map by approximating the fourth order statistics of
$\delta T_b$ as they would be for a Gaussian random field.  If this is still a
good approximation the lensing noise will be reduced. This is uncertain,
however, since during reionization the field will clearly not be Gaussian,
especially when the neutral fraction is low.  A definitive resolution of these
uncertainties will not be available until the observations are done. Here we
model fluctuations in the baryons in the same way as in
section~\ref{redshift_slices} with linear structure formation and redshift
distortions.

\begin{figure}
\centering \includegraphics[width=4.5in]{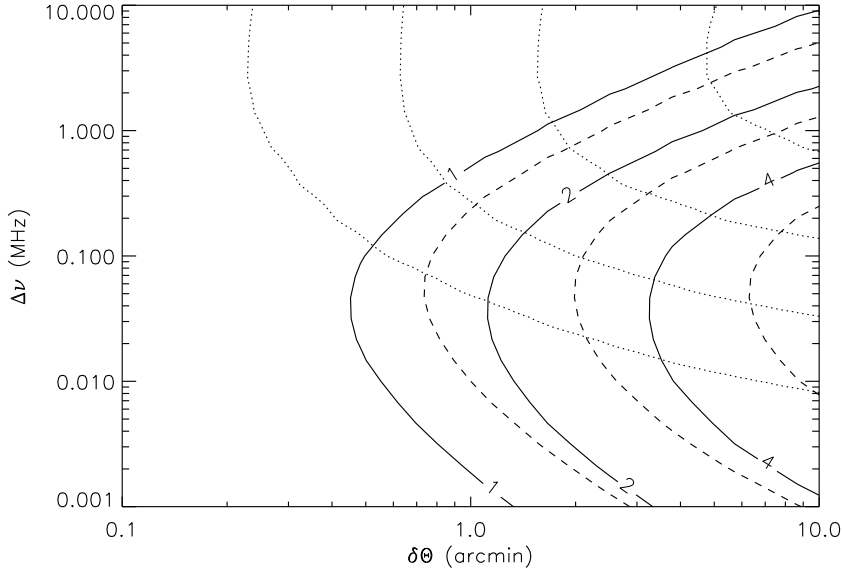}
\caption{The signal-to-noise, $\sigma_{\tilde{\kappa}}(\delta\Theta,\Delta\nu)
  /\sigma_\kappa(\delta\Theta)$, ratio for an SKA-like observation as a
  function of bandwidth and patch size.  In all cases the contours are 1, 2, 4
  and 8$\sigma$ (when visible).  The solid contours are for a covering
  fraction of $f_{\rm cover}=0.025$ and the dashed contours are for $f_{\rm
  cover}=0.018$.  
The dotted contours are the signal-to-noise ratio
  for brightness temperature with the same telescope parameters and $f_{\rm
  cover}=0.018$.  The other telescope parameters are $T_{\rm sys}=200$~K,
  $D_{\rm tel}=6$~km, $t_o=90$~days and the universe is assumed to be neutral
  in the redshift range $z=7$ to 13 in all cases.}
\label{fig:nu_theta_contour}
\end{figure}

Figure~\ref{fig:nu_theta_contour} shows the signal-to-noise ratio, defined as
$\sigma_{\tilde{\kappa}}(\delta\Theta,\Delta\nu)/\sigma_\kappa(\delta\Theta)$,
for our SKA-like telescope. For the assumptions taken
here the telescope should be able to make images ($2\sigma$) of the
dark matter on $1.3$ to 2.5~arcmin scales in 90~days ($f_{\rm cover}=0.018$ to
0.025).  These values are not too far from the optimal values and increasing the telescope's 
covering fraction or resolution would markedly improve upon these.

Unlike in the irreducible noise only case shown in figure~\ref{sigKsigG}, when
thermal noise is added the noise in $\tilde{\kappa}$ does not go to an
asymptotic value at small $\delta\Theta$.  This is because the noise increases
very rapidly near the maximum resolution of the telescope because of a
combination of effects.  First the intrinsic temperature power spectrum goes
down at $\ell \simgt 1000$.  $\overline{C}_\ell(\nu)$ is also suppressed by a
factor of $\sim 1/\Delta\nu$ when $\ell > D(z)/\delta r$ where $\delta r$ is
physical width corresponding to the bandwidth.  In addition, for a fixed
baseline the highest resolution is attained for only a limited range of
frequency which limits the number of redshift bins.  With the parameters
adopted, the cross-correlation in the temperature between frequency channels
never becomes important because the noise generally dominates when
$\Delta\nu$ is small and it is assumed to have no cross-correlation between
channels.

As can be seen from figure~\ref{fig:nu_theta_contour}, there is an
optimal bandwidth for measuring lensing.  This comes about because at
large bandwidths the number of independent frequency bins is limited.
At small bandwidth the the signal-to-noise goes down because
$C_\ell^N$ goes up faster than $\overline{C}_\ell$ with decreasing
$\Delta\nu$ for scales $\ell < D(z)/\delta r$.  This optimal bandwidth
is $\sim 0.05$~MHz for our examples.  If there is more structure on
smaller scales, such as when there are ionized bubbles, the optimal
frequency will decrease.

The optimal bandwidth for lensing is generally smaller than the optimal
bandwidth for measuring the brightness temperature itself as can also be seen
in figure~\ref{fig:nu_theta_contour}.  At the optimal bandwidth the lensing
map can have good fidelity while the temperature map is noise dominated on the
same scale.  This is a somewhat counter-intuitive situation which reflects the
fact that it is better to get more independent redshift slices at low
signal-to-noise than to image the temperature in fewer channels.  With a wider
bandwidth the temperature can be imaged on the same angular scale as the mass
distribution.  This indicates that it may be advantageous to use several
bandwidths simultaneously.

There are many additional challenges to observing 21~cm radiation from high
redshift.  The galactic foreground from synchrotron emission is about four
orders of magnitude brighter than the 21~cm signal at $\sim 180$~MHz and goes
up with decreasing frequency as $\nu^{-2.6}$.  Both this emission and
extragalactic foreground sources can, however, be cleaned from the data
because they are much smoother in frequency space (and also in position on the
sky for the Galactic foreground) than the 21~cm signal itself.  At large
frequency separations foreground emission may also decorrelate.  Generally,
foregrounds pose no more of a problem for mass-mapping than for direct mapping
of the 21~cm itself.  Rapid increases in foreground emission and in the
refractive index of the ionosphere with decreasing frequency make observations
at higher redshift progressively more difficult.  The high-redshift 21~cm
absorption ($z\simgt 30$) will be very difficult to observe and there are no
mature plans to do so at this time.  The ionosphere is opaque below $\sim
10$~MHz or $z\simgt 150$, so in principle all lower redshifts are accessible
from the ground. In practice, the high, time-dependent index of refraction
will make it difficult to go beyond $\sim 60$~MHz without major advances in
telescope technology.  The ultimate high redshift 21~cm telescope would be
located on the far side of the Moon where the absence of terrestrial
interference or an ionosphere would allow access to higher redshifts.
However, the large collecting area required would make this both technically
challenging and expensive.

Much will depend on future instrument design and the as yet unknown
characteristics of the 21~cm absorption/emission, particularly around the
epoch of reionization.  Despite this, the planned specifications for SKA may
enable it to make high-fidelity maps of the matter distribution and if enough
area can be surveyed very good statistical information should be accessible.
Realistic upgrades to the collecting area and array size would greatly
improve its ability to make mass maps.

\section{conclusion}

We have shown that when low-frequency radio telescopes become sufficiently
powerful to map the signal from high-redshift 21~cm emission/absorption within
a bandwidth of $\sim 0.05$~MHz, the data will necessarily be good enough to
map the gravitational shear due to foreground matter.  Increasing the
resolution of the telescope {\it reduces} the intrinsic noise in the shear map
both because of the number of statistically independent redshift slices
increases and because the number of independent patches on the sky increases.
As a result, 21~cm lensing offers the potential of producing high resolution,
high signal-to-noise images of the cosmic mass distribution. Such images would
be of enormous value for the study of cosmology and galaxy formation.

For the specific problem of estimating cosmological parameters the
requirements on resolution and redshift range are not particularly demanding,
but survey area is of great importance.  Even for a full-sky survey with a
pixel of radius $\delta\theta=1$~arcmin (2~arcmin fwhm) and 10\% noise per
pixel, the shear power spectrum would be cosmic variance dominated up to
$\ell \sim 10^3$.  The cosmic variance limit is probably achievable up to
$\ell\sim 10^4$ with an array of $\sim 5$~km in diameter and a covering factor
of several percent.  Cross-correlating several redshift slices with each other
and with galaxy lensing surveys over a significant portion of the sky would
begin a new era of very high precision cosmology.

The study of structure formation would benefit particularly from
higher resolution observations, however.  If a resolution of $\sim
6$~arcsec (fwhm) could be achieved, every halo more massive than the
Milky Way's would be clearly visible back to $z\sim 10$.  Even with a
resolution of $\sim 1$~arcmin (fwhm) all the halos $\simgt
2\times10^{13}\msun$ should be individually detected.  Connecting
these mass maps to images of emission at other wavelengths would
provide a tremendous wealth of information about the evolution of
structure and the formation of galaxies.

\vspace{0.5cm}

{\footnotesize RBM would like to thank B. Ciardi, P. Madau and H. Sandvik for very
useful discussions.  We would also like to thank U. Seljak and O. Zahn
for very useful comments.}


\appendix

\section{parameter estimation and visibility space}
\label{app:param-estim-visib}

The observations of 21~cm emission/absorption will be done with radio
interferometers so it is appropriate to make an explicit connection between
the observables from such an instrument and the quantities used in this
paper. The flat sky approximation greatly simplifies the mathematics and is
well justified for the angular scales of interest here.  A radio
interferometer measures the visibility, $V({\bf u})$, which in our case is
related to the spin temperature by
\begin{eqnarray}
V({\bf u}) & = &  \int d \boldsymbol{\theta}~A(\boldsymbol{\theta}) T((\boldsymbol{\theta}) e^{i 2\pi{\bf u}\cdot \boldsymbol{\theta}} \\
& = &  \int d^2{ w} \tilde{A}({\bf w}) \tilde{T}({\bf u}-{\bf w})
\end{eqnarray}
where $A(\theta)$ is the primary beam of the telescopes which is typically normalized to one at its peak ($\boldsymbol{\theta}=0$) which gives its Fourier transform, $\tilde{A}({\bf u})$, a normalization of one in ${\bf u}$-space.  Units of temperature are used here (flux density units have a factor of $2k_{\rm B}/\lambda^2$). Tildes signify Fourier transforms.  The size of the primary beam dictates the area covered on the sky in one ``pointing'' of the telescopes.  The separation of the antennas and the position on the sky dictate ${\bf u}$.

The correlation in the visibilities is given by
\begin{eqnarray}
C^V_{ij} &\equiv & \langle V^*({\bf u}_i) V({\bf u}_j)\rangle \\
& = &  \int d^2{ w}~\tilde{A}^*({\bf u}_i-{\bf w})\tilde{A}({\bf u}_j-{\bf w}) S({\bf w}) + \delta_{ij} C^N_{\bf u}\\
& \simeq & S({\bf u}_i) W_{ij} + \delta_{ij} C^N_{\bf u} \label{eq:def_W}
\end{eqnarray}
where $S({\bf u})$ 
is the intrinsic (cross-)power spectrum of the temperature and $C^N_{\bf u}$ is the noise.  It has been assumed that the temperature field is isotropic. 
Equation~(\ref{eq:def_W}) defines the window, $W_{ij}(w)$, that makes the visibilities correlated.  This in effect defines the resolution in ${\bf u}$-space.  Correlations will only exist when $\Delta {\bf u} \equiv |{\bf u}_i-{\bf u}_j|$ is less than twice the width of $\tilde{A}({\bf u})$, i.e. the smaller the telescopes are the narrower $\tilde{A}({\bf u})$ will be and the higher the resolution.  This width will be denoted $\sigma_u({\bf u})$.  It has been assumed in equation~(\ref{eq:def_W}) that the intrinsic power spectrum does not change very much on the scale of $\sigma_u$. The coordinate ${\bf u}$ is conjugate to the angle on the sky $\boldsymbol{\theta}$ so the resolution is linked to the area covered on the sky by $f_{\rm sky} \simeq (2\pi \sigma_u)^{-2}$.

The visibility power spectrum can be related to the spherical harmonic power spectrum through
\begin{eqnarray}
u^2 S(u) & = & \frac{u}{2\pi} \sum_\ell (2\ell+1) C_\ell J_{2\ell+1}(4\pi u) \\
& \simeq & \left. \frac{\ell(1+\ell)}{(2\pi)^2} C_\ell \right|_{\ell=2\pi u}
\end{eqnarray}
where the second approximation is very good for $\ell \simgt 60$ \nocite{1999ApJ...514...12W}({White} {et~al.} 1999).
If there were just one visibility measured the Fisher matrix would be 
\begin{eqnarray}
{\bf F}({\bf u})_{ab} = \frac{1}{2}{\rm tr}\left[ (C^V_{\bf u})^{-1} C^V_{\bf u},_a (C^V_{\bf u})^{-1} C^V_{\bf u},_b \right]
\end{eqnarray}
\nocite{1997ApJ...480...22T}(see {Tegmark}, {Taylor} \&  {Heavens} 1997, for a review of likelihood methods in astronomy).
Visibilities within $\sim \sigma_u$ of each other will be correlated, but an estimate of the total Fisher matrix can be made by assuming one measurement per correlated region which gives
\begin{eqnarray}
{\bf F}_{ab} & = & \sum_{\bf u} {\bf F}({\bf u})_{ab} 
\simeq  \int \frac{d^2{ u}}{\sigma_u^2} {\bf F}({\bf u})_{ab} \\
&\simeq & \frac{1}{(2\pi\sigma_u)^2} \sum_{\ell m} {\bf F}(\ell=2\pi|{\bf u}|)_{ab} \\
&\simeq &f_{\rm sky} \sum_\ell (2\ell+1) {\bf F}(\ell=2\pi u)_{ab}.
\end{eqnarray}
This is the formula~(\ref{eq:fisher}) in section~\ref{sec:darkenergy}
used to calculate cosmological parameter constraints.  It can be seen
that the $f_{\rm sky}$ factor comes from the correlations or
resolution in visibility space.  A more sophisticated treatment would
allow for partially correlated visibilities within $\sigma_u$ of each
other which would reduce the noise further.  The Fisher matrix is an
estimate of the expected inverse correlation matrix in the model
parameters at the maximum likelihood solution.  Thus $\left({\bf
    F}^{-1}\right)_{aa}$ is an estimate of the error in the parameter
$a$ after marginalizing over the other parameters.  This formalism as
outlined here in terms of the temperature, but it is equally valid for
the lensing shear or convergence.

The size of individual antennas in future radio telescope arrays are expected to be of order a few wavelengths or smaller as in the case of dipole antennas.  In this case the primary beam covers almost the whole hemisphere.
However, subtracting interference and handling the huge data rate will probably require synthesizing a much smaller beam.
In addition, the subtraction of galactic foregrounds will probably not
be possible in some regions near the galactic plane.  For these
reasons the sky fraction, $f_{\rm sky}$, and the shape of the observed
fields will be limited for a single pointing of the telescope beam.
The sky fraction and thus the $\ell$-space resolution can be increased
with multiple pointing or mosaicking.

\section{lensing in Fourier space}
\label{app:lens-four-space}

The temperature on the sky is to first order
\begin{eqnarray}
T(\boldsymbol{\theta},\nu) = {\mathcal T}(\boldsymbol{\theta},\nu) + \vec{\nabla}\Phi(\boldsymbol{\theta}) \cdot \vec{\nabla}{\mathcal T}(\boldsymbol{\theta},\nu) 
\end{eqnarray}
where ${\mathcal T}(\boldsymbol{\theta},\nu)$ is the temperature before lensing and $\Phi({\bf L})$ is the lensing potential defined so that the deflection angle is $\boldsymbol{\alpha}(\boldsymbol{\theta})=\vec{\nabla}\Phi(\boldsymbol{\theta})$. 
 The Fourier transform of this is
\begin{equation}
T(\boldsymbol{\ell},\nu) = {\mathcal T}(\boldsymbol{\ell},\nu) - \int \frac{d^2{ \ell'}}{(2\pi)^2}~ \boldsymbol{\ell}'\cdot(\boldsymbol{\ell}-\boldsymbol{\ell}') \Phi(\boldsymbol{\ell}') {\mathcal T}(\boldsymbol{\ell}-\boldsymbol{\ell}',\nu) 
\end{equation}
and, as a result, to first order
\begin{eqnarray}\label{eq:potential}
\left\langle T(\boldsymbol{\ell},\nu) T^*(\boldsymbol{\ell}-{\bf L},\nu' )\right\rangle = \left[ {\bf L}\cdot \boldsymbol{\ell}\, C_\ell(\nu,\nu') + {\bf L}\cdot ({\bf L}-\boldsymbol{\ell})\, C_{|\ell-L|}(\nu,\nu') \right] \Phi({\bf L})
\end{eqnarray}
where ${\bf L} \neq 0$ and  the average is over realizations of the temperature field while keeping the lensing potential fixed \nocite{2002ApJ...574..566H}({Hu} \& {Okamoto} 2002).  Note that if the noise is homogeneous it will drop out of this equation and the $C_\ell$'s will be only the power spectrum of the fluctuations in brightness temperature.

The observed temperature is always binned into frequency channels or bands and smoothed by the telescope's beam.  This observed temperature is 
\begin{eqnarray}
\overline{T}(\boldsymbol{\ell},\nu)& = & \int dr' q_\nu(r',\Delta\nu) \int d^2l' \tilde{A}(\boldsymbol{\ell}'-\boldsymbol{\ell})T(\boldsymbol{\ell}',\nu') \\
&=&\frac{1}{D(\nu)^2} \int d^2l' \tilde{A}(\boldsymbol{\ell}'-\boldsymbol{\ell}) \int \frac{dk}{2\pi} T_{21}\left(\frac{\boldsymbol{\ell}'}{D(\nu)},k\right)  \tilde{q}_\nu(k,\nu)  
\end{eqnarray}
where the $q_\nu(r',\Delta\nu)$ is the response function for the band centered on $\nu$, $T_{21}$ is the 3D Fourier transform of the brightness temperature and 
\begin{eqnarray}
\tilde{q}_\nu(k,\nu) &=& \int dr' q_\nu(r',\Delta\nu) e^{ikr'} \\
& \simeq &  j_o\left(\frac{k \delta r(\nu,\Delta\nu)}{2}\right)e^{ikr(\nu)} \label{nu_window}
\end{eqnarray}
with
\begin{eqnarray}
\delta r(\nu,\Delta\nu) = \frac{2}{H_o\sqrt{\Omega_m\nu_o}}\left( \sqrt{\nu+\Delta\nu/2} - \sqrt{\nu-\Delta\nu/2} \right). \label{eq:dr}
\end{eqnarray}
In (\ref{nu_window}) the response function is taken to be a boxcar shape with sharp edges at $\nu+\Delta\nu/2$ and $\nu-\Delta\nu/2$.
In (\ref{eq:dr}) the fact that the universe is matter dominated at the time of the 21~cm emission/absorption is used to express the radial distances in terms of frequency.  

 As a result of beam smearing relation (\ref{eq:potential}) will become
\begin{eqnarray}
\left\langle \overline{T}(\boldsymbol{\ell},\nu) \overline{T}^*(\boldsymbol{\ell}-{\bf L},\nu')\right\rangle = (2\pi)^2 \int d^2{ \ell}' \tilde{A}(\boldsymbol{\ell}') \tilde{A}^*(\boldsymbol{\ell}' + {\bf L}) \overline{C}_{\ell+\ell'} ~~~~~~~~~~~~~~~~~~~~~~~~~~~~~~~~~~~~~~~~~~~~~~~~~~~~~~~~~~~~~~~~~~~~~ \label{eq:alias}\\
+  \int d^2{ \ell}'' \Phi(\boldsymbol{\ell}'') \int d^2{ \ell}' \tilde{A}(\boldsymbol{\ell}') \tilde{A}^*(\boldsymbol{\ell}'-\boldsymbol{\ell}''+{\bf L}) \left\{ \left( \boldsymbol{\ell}''\cdot (\boldsymbol{\ell}+\boldsymbol{\ell}') \overline{C}_{\ell+\ell'}(\nu,\nu') - \boldsymbol{\ell}''\cdot (\boldsymbol{\ell}+\boldsymbol{\ell}'-\boldsymbol{\ell}'') \overline{C}_{\ell+\ell'-\ell''}(\nu,\nu') \right) \right\} \label{eq:smoothed}
\end{eqnarray}
where the correlations between temperature modes before lensing and beam smearing
\begin{eqnarray}\label{eq:temp_correlation}
\overline{C}_\ell(\nu,\nu') & = & \frac{1}{D(\nu)^{2}} \int dk~P_{21}\left( \sqrt{\frac{\ell^2}{D(\nu)^2} + k^2}, z(\nu) \right) \tilde{q}_\nu(k,\nu)\tilde{q}^*_\nu(k,\nu') 
\end{eqnarray}
It is assumed that $D(\nu)$ does not change significantly between frequencies that are significantly correlated.  A more rigorous derivation of (\ref{eq:temp_correlation}) is in spherical harmonic space as has been done by \nocite{2004ApJ...608..622Z}{Zaldarriaga} {et~al.} (2004), but for the scales of important here the difference is very small and (\ref{eq:temp_correlation}) is considerably easier to evaluate.

There are two effects that make (\ref{eq:alias}) and
(\ref{eq:smoothed}) different from the \nocite{2002ApJ...574..566H}{Hu} \& {Okamoto} (2002)
result (\ref{eq:potential}).  The first term~(\ref{eq:alias})
represents an aliasing effect caused by the finite size of the beam.
This will cause a false signal on scales approaching the size of the
beam or surveyed region, $L \simlt 2\pi \sigma_u$, that will need to
be subtracted.  The second term (\ref{eq:smoothed}) is a kind of
smoothing of the lensing potential over a scale of $\sim 2\pi
\sigma_u$.  In the limit of a very narrow beam, a large area in angle,
the relation (\ref{eq:potential}) is recovered except with the
frequency binned power spectra.  Thus the observations will really
measure a lensing potential that is smoothed in Fourier-space in a
rather complicated fashion.

\section{convergence estimators}
\label{app:conv-estim}

In the main text of this paper we used a real-space estimators for the
shear and convergence, $\hat{\gamma}_i(\boldsymbol{\theta})$.  We consider this the most
intuitive and instructive approach.  In the weak lensing limit the
shear map can be converted to a convergence map because they are both
related to a single lensing potential by differential operators.  This is commonly
done for galaxy lensing surveys (see \nocite{Bart&Schneid99}Bartelmann \& Schneider (2001) for a
review).  The most straightforward method is to Fourier
transform the shear map, multiply by $\ell$ dependent factors and then
transform back to a convergence map \nocite{KS93}({Kaiser} \& {Squires} 1993).  Averaging this with the
$\hat{\gamma}_3$ map would produce a convergence map with less noise
than the $\hat{\gamma}_3$ map alone.  However, the gain in noise
will not be as great as in the galaxy lensing case because in the
21~cm lensing case the Fourier modes of $\hat{\gamma}_1$ and
$\hat{\gamma}_2$ are correlated unlike in the galaxy case. 
Probably a more practical approach from a technical point of view is
to go directly from visibility space to a convergence map in real-space.

Many convergence estimators in visibility or Fourier space are possible.  Our
real-space estimators can be Fourier transformed to make a set of estimators
for the Fourier modes of shear and convergence, but the Fourier estimator of
\nocite{2002ApJ...574..566H}{Hu} \& {Okamoto} (2002) has the advantage of having the lowest noise level
for one frequency bin if the temperature distribution is Gaussian and the beam
is infinitely large (in angle).  \nocite{ZandZ2006}{Zahn} \& {Zaldarriaga} (2006) find an estimator in both
angular Fourier-space and frequency Fourier-space which is optimal with the
added assumption that the frequency Fourier modes are statistically
independent and Gaussian distributed.  The statistical independence of the
modes will break down because of binning in frequency and to a lesser extent
because of the finite range in frequency.  For this reason it is difficult to
determine how bandwidth will affect the noise in their estimator.  Instead we
choose to use the \nocite{2002ApJ...574..566H}{Hu} \& {Okamoto} (2002) estimator for each frequency band
and then weight each band.

We consider a second-order estimator for the shear or convergence of the form
\begin{eqnarray}\label{eq:gamma_tot}
\tilde{\gamma}_i({\bf L}) 
& = & \sum_\nu \int d^2\boldsymbol{\ell}~ \Gamma_i(\boldsymbol{\ell},{\bf L},\nu)  ~\overline{T}(\boldsymbol{\ell},\nu) \overline{T}^*(\boldsymbol{\ell} - {\bf L},\nu).
\end{eqnarray}
where, as in the main text, $\tilde{\gamma}_{1,2}$ are estimators for
the two components of shear and $\tilde{\gamma}_{3}$ is an estimator
for the convergence.  In this case the estimators are of the form
\begin{eqnarray}
\Gamma_i(\boldsymbol{\ell},{\bf L},\nu) = D_i({\bf L}) \omega(\nu,{\bf L}) \chi(\boldsymbol{\ell},{\bf L},\nu)
\end{eqnarray}
where
\begin{eqnarray}
D_i({\bf L}) = \left\{ 
\begin{array}{c}
\left( L_1^2 - L_2^2 \right)  \\
2 L_1 L_2    \\
|{\bf L}|^2 
\end{array}\right.
\end{eqnarray}
The function $\chi(\boldsymbol{\ell},{\bf L},\nu)$ is proportional to the  \nocite{2002ApJ...574..566H}{Hu} \& {Okamoto} (2002) estimator
\begin{eqnarray}
\chi(\boldsymbol{\ell},{\bf L},\nu) = \frac{ \left[ {\bf L}\cdot \boldsymbol{\ell}\, \overline{C}_\ell(\nu) + {\bf L}\cdot ({\bf L}-\boldsymbol{\ell})\, \overline{C}_{|\ell-L|}(\nu) \right] }{\overline{C}^T_\ell(\nu) \overline{C}^T_{|\ell-L|}(\nu)} 
\end{eqnarray}
where $\overline{C}^T_\ell(\nu)$ is the power spectrum of the actual temperature while $\overline{C}_\ell(\nu)=\overline{C}^T_\ell(\nu) + \overline{C}^N_\ell(\nu) $ is the observed power spectrum which includes noise and $\omega(\nu,{\bf L})$ is a weight that is to be determined.

In the limit of an infinitely large beam the correlation between modes is 
\begin{eqnarray} \label{eq:shear_correlation}
\left\langle \hat{\gamma}_i\left({\bf L}\right) \hat{\gamma}_j^*\left({\bf L}'\right)  \right\rangle
& = & 2(2\pi)^4 \delta\left({\bf L}-{\bf L}' \right) D_i({\bf L}) D_j({\bf L})
\sum_\nu \sum_{\nu'} \omega(\nu)\omega(\nu') \int d^2{ \ell} ~\chi\left(\boldsymbol{\ell},{\bf L}\right) \chi\left(\boldsymbol{\ell},{\bf L}\right) \overline{C}^T_\ell(\nu,\nu') \overline{C}^T_{|\ell-L|}(\nu,\nu') \\
& = & (2\pi)^2 \delta\left({\bf L}-{\bf L}' \right)  D_i({\bf L}) D_j({\bf L}) N^{ij}_{\hat{\gamma}}(L)
\end{eqnarray}
It has been assumed here that the temperature is Gaussian distributed so that
the fourth moment can be written as products of second moments.  The finite
beam will cause correlations in the noise $\langle \gamma_i({\bf L})
\gamma_i({\bf L}+\delta{\bf L})\rangle \neq 0$ when $\delta L \simlt 2\pi
\sigma_u$.  The expression for these correlations is lengthy, but easily
worked out.  In general there will be a correlation between the modes for
different components of shear, unlike in real-space.  An image can be
constructed by Fourier transforming (\ref{eq:gamma_tot}) to real-space with a
smoothing window.  The noise at a point on this image will be
\begin{eqnarray}\label{sigma_from_l}
\sigma^2_i(\delta\Theta) = \int\frac{d^2L}{(2\pi)^2} \left| \tilde{W}({\bf L},\delta\Theta) \right|^2 D_i({\bf L})^2 N_{\hat{\gamma}}(L)
\end{eqnarray}
where $\tilde{W}({\bf L},\delta\Theta)$ is the Fourier transform of
the smoothing function.  This implies that
$\sigma^2_\kappa(\delta\Theta)=2 \sigma^2_1(\delta\Theta) = 2
\sigma^2_2(\delta\Theta)$ in contrast to the noise in the
real-space estimators.

As in the real-space version, the optimal frequency weights, $\omega(\nu)$,
are complicated in general, but they simplify if we make the
approximation that each frequency bin is statistically independent.
If we minimize the diagonal entries of (\ref{eq:shear_correlation}) while
requiring that the average of (\ref{eq:gamma_tot}) with (\ref{eq:potential})
reproduce the shears and the convergence, we find
\begin{eqnarray}
\omega(\nu) = \frac{1}{2} \left[ \sum_\nu \int d^2{ \ell} \frac{\left[ {\bf L}\cdot \boldsymbol{\ell}\, \overline{C}_\ell(\nu) + {\bf L}\cdot ({\bf L}-\boldsymbol{\ell})\, \overline{C}_{|\ell-L|}(\nu) \right]^2}{\overline{C}^T_\ell(\nu) \overline{C}^T_{|\ell-L|}(\nu)} \right]^{-1}
\end{eqnarray}  
These weights can be reinserted in expression~(\ref{eq:shear_correlation}),
but now allowing for correlations between frequencies,
\begin{eqnarray}\label{N_L_corr}
N_{\hat{\gamma}}(L) 
 = \frac{(2\pi)^2}{2} \left[ \sum_\nu \int d^2\ell~ \chi(\boldsymbol{\ell},{\bf L},\nu)^2 \overline{C}^T_\ell(\nu) \overline{C}^T_{|\ell-L|}(\nu) \right]^{-2} \sum_\nu\sum_{\nu'}\int d^2{ \ell}~\chi(\boldsymbol{\ell},{\bf L},\nu) \chi(\boldsymbol{\ell},{\bf L},\nu')  \overline{C}^T_\ell(\nu,\nu')\overline{C}^T_{|\ell-L|}(\nu,\nu')
\end{eqnarray}
This can be rewritten in the suggestive form
\begin{eqnarray}\label{N_L_nocorr}
N_{\hat{\gamma}}(L) 
 = \frac{(2\pi)^2N_\nu}{2N^{\rm eff}_\nu(L)} \left[ \sum_\nu \int d^2\ell~ \frac{ \left[ {\bf L}\cdot \boldsymbol{\ell}\, \overline{C}_\ell(\nu) + {\bf L}\cdot ({\bf L}-\boldsymbol{\ell})\, \overline{C}_{|\ell-L|}(\nu) \right]^2 }{\overline{C}^T_\ell(\nu) \overline{C}^T_{|\ell-L|}(\nu)} \right]^{-1}
\end{eqnarray}
where
\begin{eqnarray}\label{Neff}
N^{\rm eff}_\nu(L) & = & N_\nu \frac{\sum_\nu \int d^2\ell~
  \chi(\boldsymbol{\ell},{\bf L},\nu)^2 \overline{C}^T_\ell(\nu)
  \overline{C}^T_{|\ell-L|}(\nu)}{\sum_\nu\sum_{\nu'}\int d^2{
    \ell}~\chi(\boldsymbol{\ell},{\bf L},\nu)
  \chi(\boldsymbol{\ell},{\bf L},\nu')
  \overline{C}^T_\ell(\nu,\nu')\overline{C}^T_{|\ell-L|}(\nu,\nu')} \\
& = & \frac{(\nu_2-\nu_1)}{\Delta\nu_L} \label{nu_corr_L}
\end{eqnarray}
which is essentially the \nocite{ZandZ2006}{Zahn} \& {Zaldarriaga} (2006) estimator except for the
$N_\nu/N^{\rm eff}_\nu(L)$ factor.  $N^{\rm eff}_\nu(L)$ is the effective
number of independent frequency bins.  Line (\ref{nu_corr_L}) is an
alternative definition of the frequency correlation length.  If the frequency
bins are uncorrelated $N^{\rm eff}_\nu(L)=N_\nu$ and $\Delta\nu_L=\Delta\nu$,
but if there are correlations between bins $N^{\rm eff}_\nu(L)<N_\nu$ and
$\Delta\nu_L>\Delta\nu$.  The irreducible noise limit discussed in
section~\ref{sec:noise} corresponds to the case where $\overline{C}_\ell(\nu)
= \overline{C}^T_\ell(\nu)$ and to infinitely narrow frequency bins.

In actual data the temperature distribution will not be Gaussian, the
foregrounds will not be perfectly subtracted which will produce spurious
correlations in frequency, there will be holes in the surveyed area caused by
point source subtraction, there will be a finite and irregular beam, and the
coverage of $u$-$v$ plane will not be complete.  All these complications
making it unclear at this time what estimator will be the best choice for real
data.

\end{document}